\documentclass[ prd,preprint,preprintnumbers,amsmath,amssymb]{revtex4}
\def\a{\alpha}
\def\b{\beta}
\def\g{\gamma}
\def\c{\gamma}
\def\d{\delta}
\def\e{\eta}
\def\h{\eta}
\def\l{\lambda}

\def\m{\mu}
\def\n{\nu}
\def\r{\rho}
\def\o{\omega}

\def\s{\sigma}

\def\t{\tau}

\def\ve{\varepsilon}
\def\x{\xi}

\def\e{\varepsilon}
\def\pa{\partial}

\def\ca{{\cal A}}
\def\cb{{\cal B}}

\def\cg{{\cal G}}

\def\cl{{\cal L}}

\def\co{{\cal O}}

\def\cw{{\cal W}}

\def\cy{{\cal Y}}

\newtheorem{prop}{Proposition}
\newtheorem{theorem}{Theorem}
\begin{document}
\preprint{ULB-TH/05-19}
\title{Parity violating vertices for spin-3 gauge fields}
\author{Nicolas Boulanger}
 \altaffiliation{F.N.R.S. Postdoctoral Researcher (Belgium)}
 \email{nicolas.boulanger@umh.ac.be}
\affiliation{
Universit\'e de Mons-Hainaut, Acad\'emie Wallonie-Bruxelles, 
M\'ecanique et Gravitation, 
Avenue du Champ de Mars 6, B-7000 Mons (Belgium)
}%
\author{Sandrine Cnockaert}%
 \altaffiliation{F.N.R.S. Research Fellow (Belgium)}
 \email{scnockae@ulb.ac.be}
\affiliation{%
Physique Th\'eorique et Math\'ematique, Universit\'e Libre
de Bruxelles \\
and International Solvay Institutes, \\
U.L.B. Campus Plaine, C.P. 231, B-1050, Bruxelles (Belgium)
}%
\author{Serge Leclercq}
 \email{serge.leclercq@umh.ac.be}
\affiliation{
Universit\'e de Mons-Hainaut, Acad\'emie Wallonie-Bruxelles, 
M\'ecanique et Gravitation,  
Avenue du Champ de Mars 6, B-7000 Mons (Belgium)
}%

\date{\today}

\begin{abstract}
The problem of constructing consistent parity-violating interactions for spin-3 gauge 
fields is considered in Minkowski space. Under the assumptions of locality, 
Poincar\'e invariance and parity non-invariance, we classify all the nontrivial perturbative deformations of the Abelian gauge algebra. 
In space-time dimensions $n=3$ and $n=5$, deformations of the free 
theory  are obtained which  make the gauge algebra non-Abelian and give rise to 
nontrivial cubic vertices in the Lagrangian, at first order in the deformation parameter $g$.
At second order in $g$, consistency conditions are obtained which the five-dimensional 
vertex obeys, but which rule out the $n=3$ candidate. 
Moreover, in the five-dimensional first-order deformation case, the gauge transformations are
modified by a new term which involves the second de Wit--Freedman 
connection in a simple and suggestive way.

\end{abstract}

\maketitle

\section{\label{sec:Introduction}Introduction}

One of the most intriguing open questions in classical field theory is the 
construction of 
consistent interactions among massless fields of spin (or helicity) $s$ greater than $2$.
It is generally believed that, when perturbatively deforming Fronsdal's massless spin-$s$ 
quadratic Lagrangian \cite{Fronsdal:1978rb}, the only first-order vertex is cubic and contains 
$s$ derivatives. This is what comes out of manifestly covariant analyses 
\cite{Berends:1984wp,Berends:1984rq,Bengtsson:1985iw,Bengtsson:1983bp,Bengtsson:1986bz} and 
a light-front approach \cite{Bengtsson:1983pd}. 
Moreover, in the aforementioned works, it is found that the massless spin-3 field  
(more generally, the odd-spin fields) carries 
a color, the fields taking value in an internal anticommutative, invariant-normed algebra. This implies that there must be (self-interacting) \emph{multiplets} of spin-3 fields, 
analogously to what happens in Yang-Mills theories for spin-1 gauge fields.

In a recent paper \cite{Bekaert:2005jf}, the problem of introducing consistent interactions among spin-3
gauge fields  has been carefully analyzed in Minkowski space-time $\mathbb{R}^{n-1,1}$  
($n>3$) using BRST-cohomological methods. 
Under the assumptions of locality, Poincar\'e and parity invariance, all the   
perturbative, consistent deformations of the Abelian gauge algebra have been determined, together 
with the corresponding deformations of the quadratic action, at first order in the
deformation parameter. 
On top of the covariant cubic vertex of \cite{Berends:1984wp}, a new 
cubic vertex is found which  also corresponds to a non-Abelian gauge algebra related to an  internal, non-commutative, invariant-normed algebra (like in Yang-Mills's theories). 
This new cubic vertex brings in five derivatives of the field: it is of the form
$\cl_1\sim g_{[abc]}(h^a \pa^2 h^b \pa^3 h^c + h^a \pa h^b \pa^4 h^c)$. 

In the present paper, we determine the non-Abelian consistent deformations of the free spin-3 
gauge theory in Minkowski space-time $\mathbb{R}^{n-1,1}$, relaxing the parity invariance constraint of \cite{Bekaert:2005jf}. 
In other words, we look for all non-Abelian deformations of the free theory such that the corresponding first-order vertices involve the completely 
antisymmetric Levi-Civita density $\ve^{\m_1\ldots\m_n}$, like in Chern--Simons theories. 
As we show explicitly in Section \ref{sec:def}, such first-order parity-breaking nontrivial 
deformations exist only in space-time dimensions $n=3$ and $n=5$.  

Following the cohomological procedure of \cite{Bekaert:2005jf}, we first classify all the 
possible first-order deformations  of the spin-3 gauge algebra that contain one Levi-Civita 
antisymmetric density  (these are the ``$a_2$ terms'' in the notation of 
\cite{Bekaert:2005jf}). 
We find two such deformations  that make the algebra non-Abelian, in dimensions $n=3$ and 
$n=5\,$.   
Then, we investigate whether these algebra-deforming terms give rise to consistent first-order 
vertices. Very interestingly, both of the algebra-deforming terms do
lead to nontrivial deformations of the quadratic Lagrangian. 
The first one is defined in dimension $n=3$, for spin-3 gauge  fields that take value in an 
internal, anticommutative, invariant-normed algebra $\ca$, while the second one is defined in 
a space-time of dimension $n=5$ for fields that take value in a commutative, invariant-normed 
internal algebra $\cb$. 
However, as we demonstrate, consistency conditions at second order in $\bar{\m}$ imply that
the algebras $\ca$ and $\cb$ must also be nilpotent of order three and associative, 
respectively. 
In turn, this means that the $n=3$ parity-breaking deformation is trivial while 
the algebra $\cb$ is a direct sum of one-dimensional ideals --- 
provided the metrics which define the norms in $\ca$ and $\cb$ are positive-definite, which is 
demanded by the positivity of energy.       
Essentially, this signifies that we may consider only one \emph{single} self-interacting 
spin-3 gauge field in the $n=5$ case, similarly to what happens in 
Einstein gravity \cite{Boulanger:2000rq}. 

The paper is organized as follows. In Section \ref{sec:FreeTheory}, we 
set up our conventions and summarize the structure of the free spin-3 gauge theory we 
are going to deform. 
We comment on the special $n=3$ case which was not investigated in \cite{Bekaert:2005jf}. 
The BRST approach \cite{Barnich:1993vg,Henneaux:1998i} to the problem of consistent 
deformation is briefly recalled in Section \ref{BRSTformulation}, where our principal 
assumptions are spelled out. We then go through the main results of \cite{Bekaert:2005jf} and 
adapt them to $n=3$  when needed; this 
gives us the opportunity to recall some results and notations needed in the following 
section where the consistent deformation problem is solved. 
Section \ref{sec:def} indeed rests upon Section \ref{BRSTformulation}'s results. 
We first determine the parity-breaking algebra-deforming 
terms.  Then we proceed to determine the corresponding first-order deformations of the free Lagrangian and study some constraints that arise at second order. 
The cases $n=3$ and $n=5$ are treated separately. 
In Section \ref{sec:concl} we summarize our results and comment on further researches. 
Finally, the Appendix gathers together several technical computations 
related to Schouten identities.     

\section{Free theory}
\label{sec:FreeTheory}

The local action for a collection
$\{h^a_{\m\n\r}\}$ of $N$ non-interacting completely symmetric
massless spin-3 gauge fields in flat space-time is
\cite{Fronsdal:1978rb}
\begin{eqnarray}
S_0[h^a_{\m\n\r}] &= \sum_{a=1}^N \int 
              &  \big[ -\frac{1}{2}\,\pa_{\s}h^a_{\m\n\r}\pa^{\s}h^{a\m\n\r} +    \frac{3}{2}\,\pa^{\m}h^a_{\m\r\s}\pa_{\n}h^{a\n\r\s} \nonumber\\
&&  +
       \frac{3}{2}\,\pa_{\m}h^a_{\n}\pa^{\m}h^{a\n}
+  \frac{3}{4}\,\pa_{\m}h^{a\m}\pa_{\n}h^{a\n} -
                        3 \,\pa_{\m}h^a_{\n}\pa_{\r}h^{a\r\m\n} \big]d^n x,
\label{freeaction}
\end{eqnarray}
where $h^a_{\m}:=\eta^{\n\r}h^a_{\m\n\r}\,$.
The Latin indices are internal indices taking $N$ values. They are raised
and lowered with the Kronecker delta $\d^{ab}\,$.
The Greek space-time indices are raised and lowered with the ``mostly plus'' Minkowskian metric $\eta_{\m\n}\,$. The free action (\ref{freeaction}) is invariant under the linear gauge
transformations $\delta_{\l}h^a_{\m\n\r} = 3 \,\pa^{}_{(\m}\l^a_{\n\r)}\,$, where 
$\eta^{\m\n}\l^a_{\m\n}\equiv 0\,$. The gauge transformations are Abelian and irreducible.
Curved (resp. square) brackets on indices denote strength-one complete 
symmetrization (resp. antisymmetrization) of the indices.

The field equations read
$\frac{\d S_0}{\d h^a_{\m\n\r}} \equiv G^{\m\n\r}_a \approx 0\,$,
where
$G^{a}_{\m\n\r}:=F^{a}_{\m\n\r} -\frac{3}{2}\eta^{}_{(\m\n}F^a_{\r)}$
and $F^{a}_{\m\n\r}$ is the Fronsdal gauge-invariant tensor
$F^{a}_{\m\n\r} :=\Box h^{a}_{\m\n\r} - 3 \,\pa^{\s}\pa^{}_{(\m}h^a_{\n\r)\s} + 3 \, \pa^{}_{(\m}\pa^{}_{\n}h_{\r)}^a\,$. We denote $F_{\m}=\h^{\n\r}F_{\m\n\r}\,$. 

The gauge symmetries enable one to get rid of some components of
$h^a_{\m\n\r}\,$, leaving it on-shell with $N^n_3$ independent
physical components, where \cite{deWit:1979pe} $N^n_3=\frac{n^3-3n^2-4n+12}{6}\,$. 
Notice that there is no propagating physical degree of freedom in $n=3\,$. 

An important object is the curvature gauge-invariant tensor
\cite{Weinberg:1965rz,deWit:1979pe,Damour:1987vm}
\begin{eqnarray}
        K^a_{\a\m|\b\n|\g\r}:= 8 \pa^{}_{[\g} \pa^{}_{[\b}\pa^{}_{[\a}h^a_{\m]\n]\r]}\,,
\end{eqnarray}
antisymmetric in the three pairs $\a\m\,$, $\b\n\,$ and $\g\r\,$. 
Actually, the spin-3 curvature is invariant under gauge transformations where 
the parameters $\lambda^a_{\m\n}$ are {\emph{not}} constrained to be traceless.

Its importance, apart from gauge invariance with unconstrained gauge parameters, stems from the fact that the field equations $G^a_{\m\n\r}\approx 0$ in Fronsdal's constrained approach 
are dynamically equivalent \cite{Bekaert:2003az,Bekaert:2003zq} to the following 
field equation in the unconstrained approach:
\begin{eqnarray}
        \eta^{\a\b} K^a_{\a\m|\b\n|\g\r} \approx 0\,.
\label{TrK}
\end{eqnarray}
There exists another field equation for completely symmetric gauge fields in the 
unconstrained approach, which also involves the curvature tensor but is 
non-local \cite{Francia:2002aa} (see also \cite{Francia:2002pt}). 
The equivalence between both unconstrained field equations was proved in 
\cite{Bekaert:2003az}.
One of the advantages of the non-local field equation of \cite{Francia:2002aa} is 
that it can be derived from an action principle. 
The equation (\ref{TrK}) is obtained from the general 
$n$-dimensional bosonic mixed symmetry case \cite{Bekaert:2003az} by specifying 
to a completely symmetric rank-3 gauge field and is \cite{Bekaert:2003zq} a 
generalization of Bargmann-Wigner's equations in $n=4$ \cite{Bargmann:1948ck}. 
However, it cannot be directly obtained from an action principle. 
For a recent work in direct relation to \cite{Francia:2002aa,Francia:2002pt}, 
see \cite{Francia:2005bu}.
 
Notice that when $n=3$, the equation (\ref{TrK}) implies that the curvature 
vanishes on-shell, which reflects the ``topological'' nature of the theory in the 
corresponding dimension. 
This is similar to what happens in 3-dimensional Einstein gravity, where the vacuum 
field equations $R_{\m\n}:=R^{\a}_{~\,\m\a\n}\approx 0$ imply that the Riemann tensor 
$R^{\a}_{~\,\m\b\n}$ is zero on-shell. 
The latter property derives from the fact that the conformally-invariant Weyl tensor 
identically vanishes in dimension $3\,$, allowing the Riemann tensor to be expressed 
entirely in terms of the Ricci tensor $R_{\m\n}\,$. 
Those properties are a consequence of a general theorem (see \cite{Hamermesch} p. 394) 
which states that a tensor transforming in an irreducible representation of $O(n)$ identically vanishes if the corresponding Young diagram is such that the sum of the lengths of the first two columns exceeds $n\,$.    

Accordingly, in dimension $n=3$ the curvature tensor $K^a_{\a\m|\b\n|\g\r}$ can be 
written \cite{Damour:1987vm} as 
\begin{eqnarray}
        K^a_{\a\m|\b\n|\g\r} \equiv \frac{4}{3}( S^a_{\a\m|[\b[\g}\h_{\r]\n]} + S^a_{\b\n|[\g[\a}
 \h_{\m]\r]}
+ S^a_{\g\r|[\a[\b} \h_{\n]\m]} )\,,
\label{curvD3}
\end{eqnarray}
where the tensor $S^a_{\a\m|\n\r}$ is defined, in dimension $n=3$, by
 \begin{eqnarray}
S^a_{\a\m|\n\r}=2\pa_{[\a} F^a_{\m]\n\r}-\frac{3}{2}\big[ 2\pa_{[\a} F_{\m]}^{a} 
 \,\h_{\n\r} - \pa_{\r} F_{[\a}^{a}\,\h_{\m]\n} 
- \pa_{\n} F_{[\a}^{a}\,\h_{\m]\r} 
 +  \pa_{\a} F_{(\n}^{a}\,\h_{\r)\m} - \pa_{\m} F_{(\n}^{a}\,\h_{\r)\a}\big]\,.
\nonumber \end{eqnarray}
It is antisymmetric in its first two indices and symmetric in its last two indices. 
For the expression of $S^a_{\a\m|\n\r}$ in arbitrary dimension
$n\geqslant 1\,$, see \cite{Damour:1987vm} where the curvature tensor 
$K^a_{\a\m|\b\n|\g\r}$ is decomposed under the (pseudo-)orthogonal group $O(n-1,1)\,$. 
The latter reference contains a very careful analysis of the structure 
of Fronsdal's spin-3 gauge theory, as well as an interesting ``topologically massive''  
spin-3 theory in dimension $n=3\,$.   
%
\section{BRST formulation}
\label{BRSTformulation}
%
In the present section we summarize the BRST-cohomological procedure 
\cite{Barnich:1993vg,Henneaux:1998i} 
where consistent couplings define deformations of the solution of the master equation. 
 For more details, we refer in particular to \cite{Bekaert:2005jf} where the BRST approach is applied to the 
spin-3 gauge theory at hand. 

\subsection{Basic assumptions}

We assume, as in the traditional perturbative Noether deformation procedure,
that the deformed action can be expressed as a power series in a
coupling constant $g\,$, the zeroth-order term in the expansion
describing the free theory: $S=S_0+g\,S_1+\co(g^2)\,$.

We require that the deformed Lagrangian be invariant under the
Poincar\'e group, but explicitly breaks parity symmetry by the presence of 
a completely antisymmetric Levi-Civita density $\ve^{\m_1\ldots\m_n}$ in the deformed 
Lagrangian. 

We reject {\em trivial} deformations
arising from field-redefinitions that reduce to the identity at
order $g^0\,$ and compute only consistent deformations, in the sense 
that the deformed theory
possesses the same number of (possibly deformed) independent gauge
symmetries, reducibility identities, {\it etc.}, as the system we
started with. Finally, crucial in the cohomological approach 
\cite{Barnich:1993vg,Henneaux:1998i} is the locality requirement: the deformed 
action $S[\phi]=S_{0}[\phi]+g S_{1}[\phi]+\ldots$ must be a {\em local}
functional. The deformations of the gauge transformations, {\it etc.}, 
must be local functions, as well as the allowed field redefinitions.

A local function of some set of fields
$\phi^i$ is a smooth function of the fields $\phi^i$ and
their derivatives $\partial\phi^i$, $\partial^2\phi^i$, ...
up to some {\it finite} order, say $k$, in the number of
derivatives. Such a set of variables $\phi^i$,
$\partial\phi^i$, ..., $\partial^k\phi^i$ will be
collectively denoted by $[\phi^i]$. Therefore, a local function
of $\phi^i$ is denoted by $f([\phi^i])$.  A local $p$-form
$(0\leqslant p \leqslant n)$ is a differential $p$-form the
components of which are local functions.
A local functional is the integral of a local $n$-form.

\subsection{BRST spectrum and differential}
\label{BRSTspectrum}

According to the general rules of the BRST-antifield formalism, for the spin-3 gauge 
theory under consideration, the spectrum of fields (including ghosts) and antifields together with their respective ghost and antighost numbers is given by \cite{Bekaert:2005jf}
\begin{itemize}
\item the fields $h^a_{\m\n\r}\,$, with ghost number $0$ and antighost number $0$;
\item the ghosts $C^a_{\m\n}\,$, with ghost number $1$ and antighost number $0$;
\item the antifields $h^{*\m\n\r}_a\,$, with ghost number $-1$ and antighost number $1$;
\item the antifields $C^{*\m\n}_a\,$, with ghost number $-2$ and antighost number $2\,$.
\end{itemize}
The Grassmannian parity of the (anti)fields is given by their ghost number modulo two.

The BRST-differential $s$ of  the free theory is generated by the functional
$$W_0 = S_0 [h^a] \;+ \int  ( 3\, h^{*\m\n\r}_a \, \pa_{\m} C_{\n\r}^a )\,d^nx.$$
More precisely, $W_0$ is the generator of the BRST-differential $s$ of the free theory through
$s A = (W_0, A)_{a.b.}\,$,
where the antibracket $(~,~)_{a.b.}$ is defined by
$(A,B)_{a.b.}=\frac{\d^R A}{\d \Phi^I}\frac{\d^L B}{\d \Phi^*_I} -
 \frac{\d^R A}{\d \Phi^*_I}\frac{\d^L B}{\d \Phi^I}\,$.
The functional $W_0$ is a solution of the \emph{master equation}
        $(W_0,W_0)_{a.b.}=0\,$.

In the context of the free spin-3 gauge theory, the BRST-differential $s$ decomposes into $s=\d + \g \,$.
The Koszul-Tate differential $\d$ decreases the antighost  number by one unit, while 
the first piece $\g\,$, the differential along the gauge orbits, leaves it unchanged.
 Both $\d$ and $\g$, and consequently the differential $s$, increase the ghost number by one unit.

The action of the differentials  $\delta$ and $\gamma$ gives zero on all the
fields of the formalism except in the few following cases:
$$\d h^{*\m\n\r}_a =G^{\m\n\r}_a\,,\quad
 \d C^{*\m\n}_a=-3(\pa_{\r}h^{*\m\n\r}_a-\frac{1}{n}\eta^{\m\n}\pa_{\r}h^{*\r}_a )\,,
\quad
\gamma h^a_{\m\n\r} = 3\,\pa^{}_{(\m}C_{\n\r)}^a\,.      
$$
%
\subsection{BRST deformation}
\label{deformation}
%
As shown in \cite{Barnich:1993vg}, the usual Noether procedure can be
reformulated within a BRST-cohomological framework. Any
consistent deformation of the gauge theory corresponds to a
solution $W=W_0+g W_1+g^2W_2+\co(g^3)$ of the deformed master
equation $(W,W)_{a.b.}=0\,$. Consequently, the first-order nontrivial
consistent local deformations $W_1=\int a^{n,\,0}$ are in
one-to-one correspondence with elements of the cohomology
$H^{n,\,0}(s \vert\, d)$ of the zeroth order BRST-differential
$s=(W_0,\cdot)$ modulo the total derivative $d\,$, in maximum
form-degree $n$ and in ghost number $0\,$. That is, one must
compute the general solution of the cocycle condition
\begin{eqnarray}
        s a^{n,\,0} + db^{n-1,1} =0\,,
        \label{coc}
\end{eqnarray}
where $a^{n,\,0}$ is a topform of ghost number zero and
$b^{n-1,1}$ a $(n-1)$-form of ghost number one, with the
understanding that two solutions of (\ref{coc}) that differ by a
trivial solution should be identified
        $a^{n,\,0}\sim a^{n,\,0} + s p^{n,-1}  + dq^{n-1,\,0}$ 
as they define the same interactions up to field redefinitions.
The cocycles and coboundaries $a,b,p,q,\ldots\,$ are local forms of
the field variables (including ghosts and antifields).

The corresponding second-order interactions $W_2$ must satisfy the consistency condition  $s  W_2=-\frac{1}{2} (W_1,W_1)_{a.b.}\,$ \cite{Barnich:1993vg}. 
This condition is controlled by the local BRST cohomology group $H^{n,1}(s\vert d)$.
%
\subsection{Cohomology of $\g$}
\label{cohogamma}
%

The groups
$H^*(\g)$ have recently been calculated \cite{Bekaert:2005ka} for massless spin-$s$ gauge fields represented by completely symmetric
(and double traceless when $s>3$) rank $s$ tensors.
In the special case
$s=3$, the result reads:
\begin{prop}\label{Hgamma} 
The cohomology of $\g$ is
isomorphic to the space of functions depending on
\begin{itemize}
  \item the antifields $h^{*\m\n\r}_a$, $C^{*\m\n}_a$ and their derivatives, denoted by
  $[\Phi^{*i}]\,$,
  \item the curvature and its derivatives $[K^a_{\a\m|\b\n|\g\r}]\,$,
  \item the symmetrized derivatives $\pa^{}_{(\a_1}\ldots\pa^{}_{\a_k}F^a_{\m\n\r)}$ of the Fronsdal tensor,
  \item the ghosts $C_{\m\n}^a$ and the traceless parts of $\pa^{}_{[\a}C_{\m]\n}^a$ and
  $\pa^{}_{[\a}C_{\m][\n,\b]}^a$.
\end{itemize} 
Thus, identifying with zero any $\g$-exact term in $H(\g)$, we have
$\g f=0 $
if and only if $f=
f\left([\Phi^{*i}],[K^a_{\a\m|\b\n|\g\r}],\{F^a_{\m\n\r}\},
                               C_{\m\n}^a, \widehat{T}^a_{\a\m\vert\n}, \widehat{U}^a_{\a\m\vert\b\n}
     \right)$
where $\{F^a_{\m\n\r}\}$ stands for the completely symmetrized
derivatives $\pa^{}_{(\a_1}\ldots\pa^{}_{\a_k}F^a_{\m\n\r)}$ of
the Fronsdal tensor, while $\widehat{T}^a_{\a\m\vert\n}$ denotes
the traceless part of $T^a_{\a\m\vert\n}:=\pa^{}_{[\a}C_{\m]\n}^a$ and $\widehat{U}^a_{\a\m\vert\b\n}$ the
traceless part of ${U}^a_{\a\m\vert\b\n}:=\pa^{}_{[\a}C_{\m][\n,\b]}^a\,$.
\end{prop}

 Let us introduce some useful standard notations and make some remarks.

Let $\{\o^I\}$ be a basis of the space of polynomials in the
$C_{\m\n}^a$, $\widehat{T}^a_{\a\m\vert\n}$ and $\widehat{U}^a_{\a\m\vert\b\n}$
(since these variables anticommute, this space is finite-dimensional).
If a local form $a$ is $\gamma$-closed, we have
\begin{eqnarray}
        \g a = 0 \quad\Rightarrow\quad
 a =\a_J([\Phi^{i*}],[K],\{F\})
        \o^J(C_{\m\n}^a,\widehat{T}^a_{\a\m\vert\n},\widehat{U}^a_{\a\m\vert\b\n}) + \g b\,,
\nonumber
\end{eqnarray}
If $a$ has a fixed, finite ghost number, then $a$ can only contain
a finite number of antifields. Moreover, since the
{\textit{local}} form $a$ possesses a finite number of
derivatives, we find that the $\a_J$ are polynomials. Such a
polynomial $\a_J([\Phi^{i*}],[K],\{F\})$ will be called  an
{\textit{invariant polynomial}}.
\vspace{.3cm}

\noindent {\textbf{Remark 1}:} 
 The Damour-Deser identity \cite{Damour:1987vm} $\eta^{\a\b}K_{\a\m|\b\n|\g\r}\equiv 2\, \pa_{[\g}F_{\r]\m\n}\,$ implies that the derivatives of the
Fronsdal tensor are not all independent  of the curvature tensor
$K$. Therefore only the completely
symmetrized derivatives of $F$ appear in Proposition \ref{Hgamma}, while the
derivatives of the curvature $K$ are not restricted. However, in the sequel we will
assume that every time the trace $\eta^{\a\b}K_{\a\m|\b\n|\g\r}$
appears, we substitute $2\pa_{[\g}F_{\r]\m\n}$ for it. We can then write
$\a_J([\Phi^{i*}],[K],[F])$ instead of the unconvenient notation
$\a_J([\Phi^{i*}],[K],\{F\})$.

\vspace{.3cm}

\noindent {\textbf{Remark 2}:} 
Proposition \ref{Hgamma} must be slightly modified in the special $n=3$ case. 
As we said in the Introduction, the curvature tensor $K$ can be expressed in terms of 
the first partial derivatives of the Fronsdal tensor, see (\ref{curvD3}).  
Moreover, the ghost variable $\widehat{U}^a_{\a\m\vert\b\n}$ identically vanishes 
because it possesses the symmetry of the Weyl tensor. 
Thus, in dimension $n=3$ we have 
\begin{eqnarray}
        \g a = 0 \quad\Rightarrow\quad a \,=\,
        \a_J([\Phi^{i*}],[F])\,
        \o^J(C_{\m\n}^a,\widehat{T}^a_{\a\m\vert\n}) + \g b\,.
\end{eqnarray} 
Another simplifying property in $n=3$ is that the variable $\widehat{T}^a_{\a\m\vert\n}$
can be replaced by its dual 
\begin{eqnarray}
\widetilde{T}^a_{\a\b}:=\ve^{\m\n}_{~~\a}\widehat{T}^a_{\m\n\vert\b} 
\quad (\widehat{T}^a_{\m\n\vert\r}=-\frac{1}{2}\ve_{\m\n}^{~~\a}\widetilde{T}^a_{\a\r})
\label{tdual} 
\end{eqnarray}
which is readily seen to be symmetric and traceless, as a consequence of the symmetries 
of $\widehat{T}^a_{\a\m\vert\n}\,$;  
\begin{eqnarray}
        \widetilde{T}^a_{\a\b}=\widetilde{T}^a_{\b\a}\,, \quad
\h^{\a\b}\widetilde{T}^a_{\a\b}= 0\,.
\label{proptdual}
\end{eqnarray}
We now remind \cite{Bekaert:2005jf} the definition of a differential which plays an important role 
in the classification of the consistent first-order deformations of $W_0\,$. 

\noindent \textbf{ Definition (differential $D$)}: The action of the differential $D$ on
$h^a_{\mu \nu\rho}$, $h^{*\mu \nu\r}_a$, $C^{*\m\n}_a$ and all
their derivatives is the same as the action of the total
derivative $d$, but its action on the ghosts is given by :
\begin{eqnarray}
D C^a_{\m\n} &=& {\frac {4}{3}} \, d x^{\a}\, {\widehat{T}}^a_{\a(\mu\vert\nu )}\,,
\nonumber \\
D  {\widehat{T}}^a_{\m\a\vert\b} &=& d x^{\r} \, {\widehat{U}}^a_{\m\a\vert\r\b}\,,
\nonumber \\
D(\partial _{( \rho } C^a_{\mu \nu)}) &=& 0\,,
\nonumber \\
D(\partial _{\rho _1 \ldots \rho _t} C^a_{\mu\nu}) &=& 0 ~ {\rm \ if \ }~ t\geqslant 2 .
\label{Ddiff}
\end{eqnarray}
The operator $D$  coincides with $d$ up to $\gamma$-exact terms.
It follows from the definitions that $D\omega^J = A^J{}_I
\omega^I$ for some constant matrix $A^J{}_I$ that involves $dx^\m$
only. 

The differential $D$ is associated with a grading.
\vspace{2mm}

\noindent \textbf{ Definition ($D$-degree)}: The number of
${\widehat{T}}_{\a\m|\n}$'s plus twice the number of
${\widehat{U}}_{\a\m|\b\n}$'s is called the $D$-degree. It is
bounded because there is a finite number of
${\widehat{T}}_{\a\m|\n}$'s and ${\widehat{U}}_{\a\m|\b\n}$'s,
which are anticommuting.

The operator $D$ splits as the
sum of an operator $D_1$ that raises the $D$-degree by one unit,
and an operator $D_0$ that leaves it unchanged. The differential 
operator $D_0$ has the same action as $d$ on $h_{\mu \nu\r}$, $h^{*\mu \nu\r}$,
$C^{*\alpha\b}$ and all their derivatives, and gives $0$ when
acting on the ghosts. The operator $D_1$ gives $0$ when acting on all the
variables but the ghosts on which it reproduces the action of $D$.
%
\subsection{Cohomological group $H^n_2(\d| d)$ and $H^{n}_k(\d| d,H(\g))$,  
$k\geqslant 2$}
\label{cohodelta}
%
We first mention a result proved in  \cite{Bekaert:2005ka,Nazim,Barnich:2005bn} 
and needed later. 
\begin{prop}\label{H2}
A complete set of representatives of $H^n_2(\d\vert d)$ is given by the antifields
$C_a^{*\m\n}$, up to explicitly $x$-dependent terms. In other terms,
\begin{eqnarray}
        \left.
        \begin{array}{ll}
         \delta a^n_2 + d b^{n-1}_1 = 0\,,
        \nonumber \\
         \quad a^n_2 \sim  a^n_2 + \delta c^n_3 + d c^{n-1}_2
         \end{array}\right\} 
          \quad \Longleftrightarrow \quad
        \left\{ \begin{array}{ll}
        a^n_2 = L^a_{\m\n}(x)C_a^{*\m\n}d^n x + \delta b^n_3 + d b_2^{n-1}\,,
        \nonumber \\
   L^a_{\m\n}(x) = \lambda^a_{\m\n} + P^a_{\m\n}(x)\,.
\end{array}\right.
\end{eqnarray}
The tensors $\lambda^a_{\m\n}$ and $P^a_{\m\n}$ are symmetric and traceless
in the indices $(\m\n)\,$. The tensor $\lambda^a_{\m\n}$ is constant whereas the tensor $P^a_{\m\n}(x)$ depends on the coordinates $x^{\m}$ explicitly. 
\end{prop}

Then, we recall that the cohomological groups $H^n_k(\d| d)$ vanish for 
$k>2\,$, which is a consequence of the fact that the free theory at hand is linear 
and possesses no reducibility (cf. Theorem 9.1 of \cite{Barnich:1994db}).    

Finally, the following result is a cornerstone of the cohomological deformation 
procedure:
\begin{prop}\label{equivcoho}
For the free spin-3 gauge theory of Fronsdal, we have 
$H^n_k(\d |d , H(\g))\cong 0$, $k>2$, and the nontrivial elements of 
$H^n_2(\d |d , H(\g))$ are the same as for $H^n_2(\d |d)\,$.    
\end{prop}
Essentially, using Proposition \ref{H2} and $H^n_k(\d| d)\cong 0$ ($k>2$), 
the proof of Proposition \ref{equivcoho} goes as follows.  
It consists in showing that, if the invariant polynomial $a^n_k$ ($k\geqslant 2$) 
is $\d\,$-trivial modulo $d\,$, $a^n_k = \d \m_{k+1}^n + d \n^{n-1}_k\,$, then one 
can always choose $\m_{k+1}^n$ and $\n^{n-1}_k$ to be invariant.  

Proposition \ref{equivcoho} is proved in \cite{Bekaert:2005jf} for $n>3\,$. We now demonstrate that it 
also holds for $n=3\,$. 
Without recalling all the details, let us point out the place where, 
in Section 4.6.2 of \cite{Bekaert:2005jf}, 
the proof of Proposition \ref{equivcoho} must be adapted to $n=3\,$. 
It is when one makes use of the projector on the symmetries of the Weyl tensor. 
In \cite{Bekaert:2005jf} we have the equations (4.45) and (4.46) 
\begin{subequations}\label{truc}
\begin{eqnarray}
    &    {Y}^{\m\n\r}_{k+1} =
        \pa_{\a} [ T_{k+1}^{\a\m\vert\n\r}+\textstyle{\frac{1}{n-1}}\eta^{\n\r}T_{k+1}^{\a\vert\m} ]
        + \delta (...)\;
\Leftrightarrow\;
\pa_{\a}T^{\a\m|\n\r}_{k+1} = Y_{k+1}^{\m\n\r}-
\textstyle{\frac{1}{n}}\h^{\n\r}\h_{\a\b}Y_{k+1}^{\m\a\b}+\d (...)\,\quad & 
\label{truc3}\\
&T_{k+1}^{\a[\m\vert\n]\r}+\textstyle{\frac{1}{n-1}}T_{k+1}^{\a\vert[\m}\eta^{\n]\r}
=\pa_{\b}S^{\b\a\vert\m\n\vert\r}_{k+1}+\d (...) &
\label{derStoT}
\end{eqnarray}
\end{subequations}
which are instrumental in order to obtain 
\begin{eqnarray}
{Y}^{\m\n\r}_{k+1} = \pa_{\a}\pa_{\b}\pa_{\c}\Psi^{\a\m\vert\b\n\vert\c\r}_{k+1}
                              + {\cg}^{\m\n\r}{}_{\a\b\g}\widehat{X}^{\a\b\g}{}_{k+1}+\d(\ldots)\,,
\label{result1}
\end{eqnarray}
where ${Y}^{\m\n\r}_{k+1}$ is completely symmetric (the subscript denotes the antighost
number), $T_{k+1}^{\a\m\vert\n\r}=-T_{k+1}^{\m\a\vert\n\r}=-T_{k+1}^{\m\a\vert\r\n}\,$, 
$\h_{\n\r}T_{k+1}^{\a\m\vert\n\r}\equiv 0\,$, $S^{\b\a\vert\m\n\vert\r}_{k+1}$ is 
antisymmetric in $(\b\a)$ and $(\m\n)$, satisfies 
$S^{\b\a\vert[\m\n\vert\r]}_{k+1}\equiv 0\equiv\h_{\n\r}S^{\b\a\vert\m\n\vert\r}_{k+1}\,$. 
The tensor $\Psi^{\a\m\vert\b\n\vert\c\r}_{k+1}$ possesses the symmetries of the curvature 
$K^{\a\m\vert\b\n\vert\c\r}$ and ${\cg}^{\m\n\r}{}_{\a\b\g}$ is the second-order differential 
operator appearing in the equations of motion 
$0\approx G^{\m\n\r}={\cg}^{\m\n\r}{}_{\a\b\g}h^{\a\b\g}\,$.  
Finally, we have 
\begin{eqnarray}
        \widehat{X}_{\a\b\g\vert k+1} := \frac{2}{n-2}\cy^{\s\t\r}_{\a\b\c}\Big(
        -S^{\m}_{~\;\s|\m\t|\r\;k+1}
        +\frac{1}{n}\eta_{\s\t}[S_{\m\n|~~\;|\r\;k+1}^{~~~\m\n}+
        S_{\m\n|~\;\r|~~k+1}^{~~~\m~~\n}]\Big)
        \label{result2}
\end{eqnarray}
where $\cy^{\s\t\r}_{\a\b\c}=\cy^{(\s\t\r)}_{(\a\b\c)}$ projects on completely symmetric  rank-3 tensors. 
The tensors ${Y}^{\m\n\r}_{k+1}$, $T_{k+1}^{\a\m\vert\n\r}$, 
$S^{\b\a\vert\m\n\vert\r}_{k+1}$ and $\Psi^{\a\m\vert\b\n\vert\c\r}_{k+1}$ are invariant.
   
In order to obtain the important Eq.(\ref{result1}), one had to project 
$\pa_{\b}S^{\b\a\vert\m\n\vert\r}_{k+1}$
on the symmetries of the Weyl tensor \cite{Bekaert:2005jf}. 
In dimension $3$, this gives zero identically. 

If we denote by $W^{\b\vert\m\n\vert\a\r}_{k+1}$ the projection
${\cw}^{\m\;\n\;\a\;\r\;}_{\m'\n'\a'\r'}S^{\b\a'\vert\m'\n'\vert\r'}_{k+1}$ of
$S^{\b\a\vert\m\n\vert\r}_{k+1}$ on the symmetries of the Weyl tensor, we have of course 
$W^{\b\vert\m\n\vert\a\r}_{k+1}=0$ because $W^{\b\vert\m\n\vert\a\r}_{k+1}\equiv 0\,$.  
Then, obviously
\begin{eqnarray}
        0 = \frac{2}{3}\,\pa_{\a}\pa_{\b}\Big[
        W^{\m\vert\a\n\vert\b\r}_{k+1}+W^{\m\vert\a\r\vert\b\n}_{k+1}+
        W^{\n\vert\a\m\vert\b\r}_{k+1}
+W^{\n\vert\a\r\vert\b\m}_{k+1}+
        W^{\r\vert\a\m\vert\b\n}_{k+1}+W^{\r\vert\a\n\vert\b\m}_{k+1}
        \Big]\,.\nonumber 
\end{eqnarray}
Substituting for $W^{\m\vert\a\n\vert\b\r}_{k+1}$ its expression in terms of 
$S^{\a\b\vert\g\d\vert\r}_{k+1}$ and using Eqs.(\ref{truc})
we find 
$0 = {Y}^{\m\n\r}_{k+1}-{\cg}^{\m\n\r}{}_{\a\b\g}\widehat{X}^{\a\b\g}_{k+1}+\d(\ldots)\,$, 
where $\widehat{X}^{\a\b\g}_{k+1}$ is still given by (\ref{result2}). 
The result (\ref{result1}) is thus recovered except for the first $\Psi\,$-term. 
This is linked to the fact that, in $n=3$, an invariant polynomial depends on the field 
$h_{\m\n\r}$ only through the Fronsdal tensor $F^{\m\n\r}$,  
see Eq.(\ref{curvD3}). In $n=3$, the Eqs.(4.41) and (4.42) of \cite{Bekaert:2005jf} are changed accordingly. 
The proof then proceeds as in \cite{Bekaert:2005jf}, where one sets $\Psi$ to zero.  
%
\section{Parity-breaking deformations}
\label{sec:def}
%
In this section, we first compute all possible parity-breaking and Poincar\'e-invariant first-order deformations of the Abelian spin-3 gauge algebra. We find that such deformations exist in three and five dimensions.  We then proceed separately for $n=3$ and $n=5$. We analyze the corresponding first-order deformations 
of the quadratic Lagrangian and find that they both exist. Then, consistency conditions 
at second order are obtained which make the $n=3$ deformation trivial and which 
constrain the $n=5$ deformation to involve only one \emph{single} gauge field. 
%
\subsection{Deformations of the gauge algebra}
%
The following proposition is proved in \cite{Bekaert:2005jf}:
\begin{prop}\label{antigh2}
Let $a$ be a local topform that is a nontrivial solution of the
cocycle equation $sa+db=0\,$. Without loss of generality, one can assume
that the antighost decomposition of $a$ stops in antighost number
two, i.e. 
$$ a=a_0+a_1+a_2\,.\label{defdecomps}$$
\end{prop}
Furthermore, in  \cite{Bekaert:2005jf}, the most general  parity and Poincar\'e-invariant term $a_2$ is determined. Here, we allow for parity-breaking terms and prove that new terms $a_2$  appear when the number of dimensions is 3 and 5. This is summarized in the following theorem.

\begin{theorem}\label{defa2}
Let $a=a_0+a_1+a_2$ be a local topform that is a nontrivial solution of the
equation $sa+db=0\,$. 
If the last term $a_2$ is parity-breaking and Poincar\'e-invariant, then it is trivial except
in three and five dimensions. In those cases,  modulo trivial terms, it can be written respectively
\begin{eqnarray} \label{deftrois} a_2=f^a_{~[bc]} \e^{\m\n\r}C^{* \a\b}_a C^b_{\m\a}
\pa_{[\n}C^c_{\r]\vert \b}d^3x\,\end{eqnarray}
and
\begin{eqnarray} a_2 = g^a_{~(bc)}\ve^{\m\n\r\s\t}C^{*\a}_{a~\;\m}
\pa_{[\n}C_{\r]}~^{\!\!\!\!\!\!\!b\,\b} \pa_{\a[\s}C^c_{\t]\b}d^5 x\,.
\label{a2n5}
\end{eqnarray}
The structure constants $f^a_{~[bc]}$ define an internal, 
anticommutative algebra $\ca$ while the structure constants  $g^a_{~(bc)}$ define an 
internal, commutative algebra $\cb\,$.  

\end{theorem}

\noindent{\bf Proof:}
The equation $sa+db=0$, when $a=a_0+a_1+a_2$,  is equivalent to the system
\begin{eqnarray}
\g a_0+\d a_1 +d b_0=0 \,,\label{first}\\
\g a_1+\d a_2 +d b_1=0 \,, \label{second}\\
\g a_2=0\,. \label{minute} 
\end{eqnarray} 

The last equation implies that, modulo trivial terms,
$a_2=\a_I \o^I$, where $\a_I $ is an invariant polynomial and the
$\{\o^I\}$ provide a basis of the polynomials in $C_{\m\n},
\widehat{T}_{\m\n\vert\r}, \widehat{U}_{\m\n\vert \r\s}$ (see Section
\ref{BRSTformulation}). Let us stress that, as $a_2$ has ghost number
zero and antifield number two, $\o^I$ must have ghost number two.
Then, acting with $\g$ on (\ref{second}) and using the
triviality of $d$, one gets that $b_1$ should also be an element
of $H(\g)$, {\it i.e.}, modulo trivial terms, $b_1=\b_I \o^I$,
where the $\b_I$ are invariant polynomials.

We further expand $a_2$ and $b_1$ according to the $D$-degree
defined in Section \ref{cohogamma}: 
\begin{eqnarray} 
a_2=\sum_{i=0}^M a_2^i= \sum_{i=0}^M
\a_{I_i} \o^{I_i}\,, \hspace{.5cm} b_1=\sum_{i=0}^{M}
b_1^i=\sum_{i=0}^{M} \b_{I_i} \o^{I_i}\,,\nonumber 
\end{eqnarray} 
where $a_2^i$, $b_1^i$ and $\o^{I_i}$ have $D$-degree $i$. 
The equation (\ref{second}) then reads
$$\sum_i \d [  \a_{I_i} \o^{I_i}]+\sum_i D [\b_{I_i} \o^{I_i}]= \g(\ldots)\,,$$
or equivalently
$$\sum_i \d [  \a_{I_i} ]\o^{I_i}+\sum_i D_0 [\b_{I_i}] \o^{I_i}\pm\sum_i\b_{I_{i}} A^{I_i}_{I_{i+1}} \o^{I_{i+1}}= \g(\ldots)\,,$$
where $A^{I_i}_{I_{i+1}}\o^{I_{i+1}}= D \o^{I_i}$, which implies
\begin{eqnarray} \d\a_{I_i}+d\b_{I_i}\pm\b_{I_{i-1}}
A^{I_{i-1}}_{I_{i}}= 0\, \label{third}\end{eqnarray} for each $D$-degree
$i$, as the elements of the set $\{\o^I\}$ are linearly
independent nontrivial elements of $H(\g)$.

The next step is to analyze the equation (\ref{third}) for each $D$-degree. The results depend on the dimension, so we split the analysis into the cases $n=3$, $n=4$, $n=5$ and $n>5$.

\vspace{5mm}

{\bfseries{Dimension 3}}

\begin{itemize}
  \item \underline{degree zero} : In $D$-degree 0, the  equation  (\ref{third}) reads $\d\a_{I_0}+d\b_{I_0}=0$, which implies that $\a_{I_0}$ belongs to
$H_2(\d\vert d)$. 
In antifield number 2, this group has nontrivial
elements given by Proposition \ref{H2}, which are  proportional to
$C^{*\m\n}_{a}$ . The requirement of translation-invariance
restricts the coefficient of $C^{*\m\n}_{a}$  to be constant. On the other hand, in $D$-degree 0 and
ghost number 2, we have  $\o^{I_0}=C^b_{\m\r}C^{c}_{\n\s}$. To get
a parity-breaking but Lorentz-invariant $a^0_2$, a scalar quantity must be build 
by contracting $\o^{I_0}$,  $C^{*\m\n}_{a}$, the tensor $\e^{\m\n\r}$ and a product of
$\eta_{\m\n}$'s. This cannot be done because there is an odd number of indices, so $a^0_2$ vanishes: $a^0_2=0$. One can then also choose $b^0_1=0$.

  \item \underline{degree one} : We now analyze Eq.(\ref{third}) in $D$-degree 1. It reads 
  $\label{blabla} \d\a_{I_1} +d\b_{I_1}= 0\,$ and implies that $\a_{I_1} $ is an
  element of $H_2(\d\vert d)$. 
  Therefore the only parity-breaking and Poincar\'e-invariant $a^1_2$ that can be built is 
  $a^1_2=f^a_{~bc}\e^{\m\n\r} C_a^{*\a\b}C^{b}_{\a\m} T^{c}_{ \n\r\vert \b}d^3x\,$. 
  Indeed, it should have the structure $\e C^*C\widehat{T}$ (or $\e C^*CT$, up to trivial 
  terms), contracted with $\h$'s. In an equivalent way, it must have the structure 
  $C^*C\widetilde{T}\,$, contracted with $\h$'s, where the variable $\widetilde{T}$ was 
  introduced in Eq.(\ref{tdual}). Due to the symmetry properties (\ref{proptdual}) of 
  $\widetilde{T}\,$ which are the same as the symmetries of $C^a_{\m\n}$ and 
  $C^{*\m\n}_{a}$, there is only one way of contracting $\widetilde{T}\,$, 
  $C$ and $C^*$ together: $f^a_{~bc}C^{*\m\n}_{a}C_{\m}^{b\,\r}\widetilde{T}^c_{\n\r}\,$. 
  Of course, no Schouten identity  (see Appendix \ref{A}) can come into play because of the 
  number and the 
  symmetry of the fields composing 
  $f^a_{~bc}C^{*\m\n}_{a}C_{\m}^{b\,\r}\widetilde{T}^c_{\n\r}\,$. The latter term is 
  proportional to  
  $a^1_2=f^a_{~bc}\e^{\m\n\r} C_a^{*\a\b}C^{b}_{\a\m} T^{c}_{ \n\r\vert \b}d^3x\,$, up to 
  trivial terms. 
  One can now easily compute that $b^1_1=- 3\,f^a_{~bc}\e^{\m\n\r} 
  (h_a^{*\a\b\l}-\frac{1}{3}\eta^{\a\b}
  h_a^{*\l})C^{b}_{\a\m}\widehat{T}^{c}_{\n\r\vert \b}\frac{1}{2}\e_{\l\s\t}dx^\s  dx^\t\,$.

  \item \underline{degree two} : The equation (\ref{third}) in $D$-degree 2 is 
  $\d\a_{I_2}+d\b_{I_2}-\b_{I_{1}}A^{I_{1}}_{I_{2}}=0$, with
  \begin{eqnarray} 
  -\b_{I_{1}}A^{I_{1}}_{I_{2}}\o^{I_2}&=&3\,f^a_{~bc}\e^{\m\n\r} 
  (h_a^{*\a\b\l}-\frac{1}{3}\eta^{\a\b}
  h_a^{*\l}) (\frac{4}{3} \widehat{T}^{b}_{\h (\a\vert\m)} \widehat{T}^{c}_{ \n\r\vert \b}) \frac{1}{2} \e_{\l\s\t} dx^\h dx^\s  dx^\t   \nonumber \\
  &=&2\ f^a_{~(bc)}\e^{\m\n\r} 
  (h_a^{*\a\b\l}-\frac{2}{3}\eta^{\a\b}
  h_a^{*\l})  \widehat{T}^{b}_{\l \m\vert\a} \widehat{T}^{c}_{ \n\r\vert \b}  \ d^3 x   
  \,.
  \nonumber 
  \end{eqnarray}
The latter equality holds up to irrelevant trivial $\g$-exact terms. It is obtained 
by using the fact that there  are only two linearly independent scalars having the structure 
$\ve h^* \widehat{T}\widehat{T}\,$. They are 
$\ve^{\m\n\r}h^{*\a\b\g}\widehat{T}_{\m\n|\a}\widehat{T}_{\r\b|\g}$ and 
$\ve^{\m\n\r}h^{*\a}\widehat{T}_{\m\n|}^{~~\b}\widehat{T}_{\r\b|\a}\,$. 
To prove this, it is again easier to use the dual variable $\widetilde{T}$ instead of $\widehat{T}$. 
One finds that the linearly independent terms with the structure 
$\ve h^* \widetilde{T}\widetilde{T}$ are
$f^a_{~(bc)}\ve^{\m\n\r}h^{*a}_{~\m}{\widetilde{T}}_{\n}^{b\,\a}{\widetilde{T}}^c_{\r\a}$ 
and 
$f^a_{~(bc)}\ve^{\m\n\r}h^{*a\,\a\b}_{~\m}{\widetilde{T}}^{b}_{\n\a}{\widetilde{T}}^c_{\r\b}\,;$
 they are proportional to  
$f^a_{~(bc)}\ve^{\m\n\r}h_a^{*\a\b\g}\widehat{T}^b_{\m\n|\a}\widehat{T}^c_{\r\b|\g}$ and 
$f^a_{~(bc)}\ve^{\m\n\r}h^{*\a}_a\widehat{T}_{\m\n|}^{b~\;\b}\widehat{T}^c_{\r\b|\a}\,$.

Since the expression for $\b_{I_{1}}A^{I_{1}}_{I_{2}}$ is not $\d$-exact modulo $d\,,$ it must vanish: $f_{abc}=f_{a[bc]}\,.$ One then gets 
that $\a_{I_2}$ belongs to
$H_2(\d\vert d)$.  However, no such parity-breaking and Poincar\'e-invariant $a^2_2$ can be formed in $D$-degree 2, so $a_2^2=0=b_1^2\,.$ 
  
  \item \underline{degree $>2$} : Finally, there are no $a^i_2$ for $i>2$. Indeed, there is no ghost
combination $\o^{I_i}$ of ghost number two and $D$-degree higher
than two, because $\widehat{U}$ identically vanishes when $n=3$.
\end{itemize}

\vspace{2mm}
{\bfseries {Dimension 4}}

There is no nontrivial deformation of the gauge algebra in dimension 4.
\begin{itemize}

\item \underline{degree zero} : The equation (\ref{third})  reads $\d \a_{I_0}+d \b_{I_0}=0$. It implies that $\a_{I_0}$ belongs to $H_2^4{(\d|d)}$, which means that $\a_{I_0}$ is of the form $k^a_{~bc}\ve^{\m\n\r\s}C_a^{*\a\b}d^4x$ where $k^a_{~bc}$ are some constants. 
It is obvious that all contractions of $\a_{I_0}$ with two undifferentiated ghosts $C$ in a Lorentz-invariant way identically vanish. One can thus choose $a_2^0=0$ and $b_1^0=0$.

\item \underline{degree one} : The equation in $D$-degree 1 reads $\d \a_{I_1}+d \b_{I_1}=0$. The nontrivial part of $\a_{I_1}$ has the same form as in $D$-degree 0. It is however impossible to build a nontrivial Lorentz-invariant $a_2^1$ because $\o_{I_1} \sim CT$ has an odd number of indices. So $a_2^1=0$ and $b_1^1=0$.

\item \underline{degree two} : In $D$-degree 2, the equation $\d \a_{I_2}+d \b_{I_2}=0$ must be studied. Once again, one has $\a_{I_2}=k^a_{~bc}\ve^{\m\n\r\s}C_a^{*\a\b}d^4x$. 
There are two sets of $\o_{I_2}$'s : $\widehat{T}^b_{\m\n\vert \a}\widehat{T}^c_{\r\s \vert \b}$ and $C^b_{\a\b}\widehat{U}^c_{\m\n|\r\s}\,.$ A priori there are three different ways to contract the indices of terms with the structure $\ve C^*\widehat{T}\widehat{T}$, but because of Schouten identities (see Appendix\,\ref{ectt}) only two of them are independent, with some symmetry constraints for the structure functions. No Schouten identities exist for terms with the structure $\ve C^*C\widehat{U}$. The general form of $a_2^2$ is thus, modulo trivial terms, 
\begin{eqnarray}
a_2^2&=&\stackrel{(1)}{k^a}_{[bc]}\ve^{\m\n\r\s}\ C_a^{*\a\b}\
   \widehat{T}^{b}_{\m\n \vert \a}\ \widehat{T}_{\r\s \vert \b}^c d^4 x 
+\stackrel{(2)}{k^a}_{(bc)}\ve^{\m\n\r\s}\
C^{*\a}_{a~\,\m}\ \widehat{T}^{b \, ~ \b}_{ \n\r\vert}\ \widehat{T}_{\s\a\vert \b}^c d^4 x
\nonumber \\
&&+\stackrel{(3)}{k^a}_{bc}\ve^{\m\n\r\s}\
C^{*}_{a\m\a}\ C^b_{\n\b}\ \widehat{U}^{c\ \ \a\b}_{\r\s|}d^4 x\,,\nonumber
\end{eqnarray}
and $b_1^2$ is given by
\begin{eqnarray}
b_1^2&=&
-3 \ \ve^{\m\n\r\s} \Big[ ( h_a^{*\l\a\b}-\frac{1}{4} 
h_a^{*\l}\h^{\a\b})\stackrel{(1)}{k^a}_{[bc]}
    \widehat{T}^{b}_{\m\n \vert \a}\widehat{T}_{\r\s \vert \b}^c\nonumber \\
&&+(h^{*\l\a}_{a~\;\;\m}-\frac{1}{4} h_a^{*\l}\d^\a_{\m})(\stackrel{(2)}{k^a}_{(bc)}
\widehat{T}^{b \, ~ \b}_{ \n\r\vert}\ \widehat{T}_{\s\a\vert \b}^c 
+\stackrel{(3)}
{k^a}_{bc}\ C^b_{\n\b}\ \widehat{U}^{c\ \ \ \b}_{\r\s\vert \a})
\Big]\frac{1}{3!}\ve_{\l \r\s\t}dx^\r dx^\s dx^\t
\,.\nonumber\end{eqnarray}

\item \underline{degree three} : Eq.(\ref{third}) now reads $\d \a_{I_3}+d \b_{I_3}+\b_{I_2}A_{I_3}^{I_2}=0$ , with
\begin{eqnarray}
\b_{I_2}A_{I_3}^{I_2}\o^{I_3}
&=&-\displaystyle\frac{3}{2}\stackrel{(1)}{k^a}_{[bc]}\ve^{\m\n\r\s}h_a^{*\l}
 \widehat{T}^{b ~~\a}_{\m\n\vert}\widehat{U}^c_{\l\a|\r\s}d^4 x\nonumber\\
&&-3\stackrel{(2)}{k^a}_{(bc)}\ve^{\m\n\r\s}h_{a~~\m}^{*\a\l}\Big(\widehat{T}^{b~~\b}_{\n\a\vert}\widehat{U}^c_{\l\b|\r\s}-\widehat{T}^{b~~\b}_{\n\r\vert}
\widehat{U}^c_{\l\b|\s\a}\Big)d^4 x\nonumber\\
&&+4\stackrel{(3)}{k^a}_{bc}\ve^{\m\n\r\s}h_{a~~\m}^{*\a\l}\widehat{T}^b_{\l(\b\vert \n)}\widehat{U}^{c\quad\b}_{\r\s|\a}d^4 x\nonumber\\
&=&\Big( -\frac{3}{2}(\stackrel{(1)}{k^a}_{[bc]}+\stackrel{(2)}{k^a}_{(bc)})
\ve^{\b\g\r\s}h_a^{*\m}\h^{\a\n}
-(6\stackrel{(2)}{k^a}_{(bc)}+4\stackrel{(3)}{k^a}_{bc})\ve^{\m\n\l\b}
h^{*\g\r}_{a~~\l} \h^{\a\s}
\Big)
\nonumber\\
&&\hspace{9cm}\times\widehat{T}^b_{\b\g\vert \a}\widehat{U}^c_{\m\n|\r\s}d^4x\nonumber
\end{eqnarray} 
The latter equality is obtained using Schouten identities 
(see Appendix\,\ref{ehtu}). 
It is obvious that the coefficient of $\o^{I_3}=\widehat{T}^b_{\b\g\vert \a}\widehat{U}^c_{\m\n|\r\s}$ cannot be $\d$-exact modulo $d$ unless it is zero. This implies that $\stackrel{(1)}{k^a}_{[bc]}=\stackrel{(2)}{k^a}_{(bc)}=\stackrel{(3)}{k^a}_{bc}=0$.
So $a_2^2$ is trivial and can be set to zero, as well as $b_1^2$. One now has $\d \a_{I_3}+d \b_{I_3}=0$, which has  the usual solution for $\a_{I_3}$, but there is no nontrivial Lorentz-invariant $a_2^3$ because there is an odd number of indices to be contracted.

\item \underline{degree $\geq 4$} : Eq.(\ref{third}) is $\d \a_{I_4}+d \b_{I_4}=0$, thus $\a_{I_4}$ is of the form $l^a_{~bc}\ve^{\m\n\r\s}C_a^{*\a\b}d^4x$. 
There are two different ways to contract the indices : $\ve^{\m\n\r\s}C_a^{*\a\b}\widehat{U}^b_{\m\n|\a\g}\widehat{U}_{\r\s|\b}^{c\ \ \ \g}$ and $\ve^{\m\n\r\s}C^{*}_{a\m\a}\widehat{U}^b_{\n\r|\b\g}\widehat{U}_\s^{c\a|\b\g}$, but  both functions vanish because of Schouten identities (see Appendix\,\ref{ecuu}). Thus $a_2^4=0$ and $b_2^4=0$. No candidates $a^i_2$ of ghost number two exist in $D$-degree higher than four because there is no appropriate $\o^{I_i}$.

\end{itemize}

\vspace{2mm}
{\bfseries{Dimension 5}}
\begin{itemize}

\item \underline{degree zero} : In $D$-degree 0, the equation (\ref{third}) reads $\d \a_{I_0}+d\b_{I_0}=0\,,$ 
which means that $\a_{I_0}$ belongs to $H^{5}_2(\d|d)$.
However, $a^0_2$ cannot be build with such an $\a_{I_0}$ because the latter has an odd number of indices while $\o^{I_0}$  has an even one. So, $\a_{I_0}$ and
$\b_{I_0}$ can be chosen to vanish. 

\item \underline{degree one} : In $D$-degree 1, the equation becomes $\d \a_{I_1}+d\b_{I_1}=0$, so $\a_{I_1}$ belongs to $H^{5}_2(\d|d)$. However, it is impossible to build a non-vanishing Lorentz-invariant $a_2^1$ because  in a product $C^*C\widehat{T}$ there are not enough indices that can be antisymmetrised to be contracted with the Levi-Civita density.
So $\a_{I_1}$ and $\b_{I_1}$ can be set to zero.

\item \underline{degree two} : The equation (\ref{third}) reads $\d\a_{I_2}+d\b_{I_2}=0$. Once again, there is no way to build a Lorentz-invariant $a_2^2$ because of the odd number of indices. So $\a_{I_2}=0$ and $\b_{I_2}=0$.

\item \underline{degree three} : In $D$-degree 3,
the equation is $\d \a_{I_3}+d\b_{I_3}=0$, so $\a_{I_3}\in H^{5}_2(\d|d)$. This gives rise to an $a_2$ of the form "$g\ve\, C^*\widehat{T}\widehat{U}d^5 x$". There is only one nontrivial Lorentz-invariant object of this form : $a_2=g^a_{~bc}\ve^{\m\n\r\s\t}C^{*}_{a\m\a}\widehat{T}^{b}_{\n\r\vert\b}
\widehat{U}^{c\a\b\vert}_{~~~~\,\s\t}d^5 x$. It is equal to (\ref{a2n5}) modulo a $\g$-exact term.
One has 
\begin{eqnarray}
b_1^3&=&\b_{I_3}\o^{I_3}=-3g^a_{~bc}\ve^{\m\n\r\s\t}
(h^{*~~\; \l}_{a\m\a}-\frac{1}{5}\eta_{\m\a}h_a^{*\l})
 \widehat{T}^b_{\n\r\vert\b}
\widehat{U}^{c\a\b}_{\ \ \ \ |\s\t}\frac{1}{4!}\ve_{\l \g\d\h \xi}dx^\g dx^\d dx^\h 
dx^{\xi}\,.\nonumber \end{eqnarray}

\item \underline{degree four} : The equation (\ref{third}) reads 
$\d \a_{I_4}+d\b_{I_4}-\b_{I_3}A^{I_3}_{I_4}=0$, with $$\b_{I_3}A^{I_3}_{I_4}\o^{I_4}=-3g^a_{~[bc]}\ve^{\m\n\r\s\t}h^{*\a\l}_{a~~\;\m}
\widehat{U}^{b}_{\l\b|\n\r}\widehat{U}^{c\ \b}_{\ \a\ \ |\s\t}d^5 x$$ 
The coefficient of $\o^{I_4}\sim\widehat{U}\widehat{U}$ cannot be
$\d$-exact modulo $d$ unless it vanishes, which implies that
$g^a_{~bc}=g^a_{~(bc)}$. One is left with the equation $\d
\a_{I_4}+d\b_{I_4}=0$, but once again it has no
Lorentz-invariant solution because of the odd number of indices to be contracted. So $\a_{I_4}=0$ and $\b_{I_4}=0$.
\item \underline{degree higher than four}: There is again no $a^i_2$ for $i>4$, for the same reasons as in four dimensions.
\end{itemize}

{\bfseries{Dimension $\mathbf{n>5}$}}

No new $a_2$ arises because it is impossible to build a non-vanishing parity-breaking term by contracting an element of $H^5_2(\d\vert d)$, {\it i.e.}  $C^{*\m\n}$, two ghosts from the set $\{ C^{\m\n}, \widehat{T}^{\m\n\vert \r}, \widehat{U}^{\m\n\vert \r\s} \}$, an epsilon-tensor $\e^{\m_1 \ldots \m_n}$ and metrics $\h_{\m\n}$.

\vspace{2mm}
Let us finally notice that throughout this proof we have acted as if
 $\a_I$'s trivial in $H^n_2(\d\vert d)$ lead to trivial $a_2$'s. The correct statement
is that trivial $a_2$'s correspond to $\a_I$'s trivial in $H^n_2(\d\vert d,H(\g))$ (see {\it e.g.} \cite{Boulanger:2000rq} for more details). However, both statements are equivalent in this case, since both groups are isomorphic (Proposition \ref{equivcoho}).

This ends the proof of Theorem \ref{defa2}.


\subsection{Deformation in 3 dimensions}
\label{dim3peu}

In the previous section, we determined that the only nontrivial first-order deformation of the free theory in three dimensions deforms the gauge algebra by the term (\ref{deftrois}). 
We now check that this deformation can be consistently lifted and leads to a consistent first-order deformation of the Lagrangian. 
However, we then show that obstructions arise at second order, {\it i.e.} that 
one cannot construct a corresponding consistent second-order deformation unless the whole 
deformation vanishes. 

\subsubsection{First-order deformation}
A consistent first-order deformation exists if one can solve Eq.(\ref{first}) for $a_0$, where $a_1$ is obtained from Eq.(\ref{second}). 
The existence of  a solution $a_1$ to Eq.(\ref{second}) with $a_2=a_2^1$ is a consequence of the analysis of the previous section. Indeed, the $a_2$'s of Theorem \ref{defa2} are those that admit an $a_1$ in Eq.(\ref{second}).
Explicitely,  $a_1$ reads, modulo trivial terms, \begin{eqnarray}
a_1=f^a_{~[bc]}\e^{\m\n\r} \Big[ 3\, (h_a^{*\a\b\l}-\frac{1}{3} \h^{\a\b} h_a^{* \l})\, (\frac{1}{3}h^{b}_{\a\m\l}T^{c}_{ \n\r\vert \b}
+\frac{1}{2} C^{b}_{\a\m} \pa_{[\r}h^{c}_{ \n] \b\l})\nonumber \\
+\frac{1}{3}h_a^{*\l} T^b_{\l\n\vert \m}h_\r^c 
+h^*_{a\m} C^{b\,\a}_{\,\n}(-\frac{1}{2} \pa^\l h^c_{\l\a\r}+\pa_{(\a}h^c_{\r)})\Big] d^3 x\,.\nonumber 
\end{eqnarray}

On the contrary, a new condition has to be imposed on the structure function for the existence of an $a_0$ satisfying Eq.(\ref{first}).  Indeed a necessary condition for $a_0$ to  exist is that $\d_{ad}f^d_{~[bc]}=f_{[abc]}$, 
which means that the corresponding internal anticommutative algebra $\ca$ is endowed with an invariant norm. The internal metric we use is $\d_{ab}$, which is positive-definite.  

The condition is also sufficient and $a_0$ reads, modulo trivial terms,
\begin{eqnarray}
a_0=f_{[abc]}\e^{\m\n\r}\Big[
\frac{1}{4} \pa_\m h^a_{\n\a\b}\pa^\a h^{b\b}h^{c}_\r
+\frac{1}{4} \pa_\m h^a_{\n\a\b}\pa^\a h^{b\b\g\d}h^{c}_{\r\g\d}
-\frac{5}{4} \pa_\m h^a_{\n\a\b}\pa^\a h^{b\g}h^{c\b}_{\r\g}\nonumber \\
-\frac{3}{8} \pa_\m h^a_{\n}\pa^\a h^{b}_{\a}h^{c}_{\r}
+\frac{1}{4}\pa_\m h^{a\a\b}_{\n}\pa^\g h^{b}_{\g}h^{c}_{\r\a\b}
-\pa_\m h^{a}_{\n}\pa^\g h^{b}_{\a\b\g}h_{\r}^{c\a\b}\nonumber \\
+\frac{1}{2} \pa_\m h^{a}_{\n\a\b}\pa^\g h^{b}_{\a\g\d}h_{\r}^{c\b\d}
+2 \pa_\m h^{a}_{\n}\pa^\b h^{b\g}h_{\r}^{c\b\g}
-\frac{1}{4}\pa_\m h^{a}_{\n\a\b}\pa^\g h^{b\a\b\d}h_{\r\g\d}^{c}\nonumber \\
-\frac{1}{4}\pa_\m h^{a}_{\n\a\b}\pa^\g h^{b\b}h_{\r\g}^{c~\a}
-\frac{5}{8}\pa_\m h^{a}_{\n}\pa_\r h^{b\b}h_{\b}^{c}
+\frac{7}{8}\pa_\m h^{a}_{\n\a\b}\pa_\r h^{b\a\b\g}h_{\g}^{c}\nonumber \\
+\frac{1}{4}\pa_\m h^{a}_{\n\a\b}\pa_\g h_\r^{b\a\g}h^{c\b}
+\frac{1}{4}\pa_\m h^{a}_{\n}\pa^\a h_{\r\a\b}^{b}h^{c\b}
-\frac{1}{4}\pa_\m h^{a}_{\n\a\b}\pa^\g h_{\r\g\d}^{b}h^{c\a\b\d}\nonumber \\
-\frac{1}{8}\pa_\m h^{a}_{\n}\pa^\a h_{\r}^{b}h^{c}_{\a}
-\frac{1}{8}\pa_\m h^{a}_{\n\a\b}\pa^\g h_{\r}^{b\a\b}h^{c}_{\g}
\Big]d^3x\,.\nonumber 
\end{eqnarray}
To prove these statements about $a_0$, one writes the most general $a_0$ with two derivatives, that is Poincar\'e-invariant but breaks the parity symmetry. One  inserts this $a_0$ into the equation to solve, {\it i.e.} $\d a_1+\g a_0=d b_0$, and computes the $\d$ and $\g$ operations. One takes an Euler-Lagrange derivative of the equation with respect to the ghost, which removes the total derivative  $d b_0$. The equation becomes $\frac{\d}{\d C_{\a\b}}(\d a_1+\g a_0)=0$, which we multiply by $C_{\a\b}$. The terms of the equation have the structure $\e C \pa^3h h$ or $\e C \pa^2h \pa h$.
One  expresses them as linear combinations of a set of linearly independent quantities, which is not obvious as there are Schouten identities relating them (see Appendix \ref{cpahh}). One can finally solve the equation for the arbitrary coefficients in $a_0$, yielding 
the above results.

\subsubsection{Second-order deformation}
Once the first-order deformation $W_1=\int (a_0+a_1+a_2)$ of the free theory is determined, the next step 
is to investigate whether a corresponding second-order deformation $W_2$ exists. 
This second-order deformation of the master equation is constrained to obey $  s W_2 = -\frac{1}{2} (W_1,W_1)_{a.b.}\,,$ (see Section \ref{deformation}).
Expanding both sides according to the antighost number yields several conditions.
The maximal antighost number condition reads 
\begin{eqnarray}
-\frac{1}{2}(a_2,a_2)_{a.b.} = \g c_2 + \d c_3 + d f_2  
\nonumber
\end{eqnarray}
where $W_2 = \int d^3x \ (c_0+c_1+c_2+c_3)$ and $antigh(c_i)=i\,$. 
It is easy to see that the expansion of $W_2$ can be assumed to stop at antighost number  
$3$ and that $c_3$ may be assumed to be invariant. 
The calculation of $(a_2,a_2)_{a.b.}$, where 
        $a_2 = f^a_{~[bc]} \varepsilon^{\mu\nu\rho} C^{*\alpha\beta}_a \, 
        C^{b}_{\mu\alpha} \partial_{{\nu}}C^c_{\rho\beta}\,$, 
gives
\begin{eqnarray}
        (a_2,a_2)_{a.b.} &=& 2 \frac{\d^R a_2}{\d C^{*\a\b}_a}\frac{\d^L a_2}{\d C_{\a\b}^a}
  \nonumber \\
  &=&\g\mu + d\n + 2f^a_{~bc}f_{ead}\ve^{\m\n\r}\ve_{\a\l\t}
  \Big[\,
  \frac{1}{2}\,C^{*e\s\x}C_{\m}^{b\,\a}{\widehat{T}}^c_{\n\r|\s}
  {\widehat{T}}^{d\l\t|}_{~~~~~\x}+
  \frac{1}{2}\,C^{*e\s\x}C_{\m\s}^{b}{\widehat{T}}_{\n\r|}^{c~~\a}
  {\widehat{T}}^{d\l\t|}_{~~~~~\x}
  \nonumber \\
  &&-\frac{1}{3}\,C^{*e\a\x}C_{\m}^{b\,\s}{\widehat{T}}^c_{\n\r|\s}
  {\widehat{T}}^{d\l\t|}_{~~~~~\x}-
  \frac{2}{3}\,C^{*e\s\x}{\widehat{T}}^{b\l}_{~~(\m|\a)}
  {\widehat{T}}^c_{\n\r|\s}C_{\x}^{d\,\t}-
  \frac{2}{3}\,C^{*e\s\x}{\widehat{T}}^{b\l}_{~~(\m|\s)}
  {\widehat{T}}^{c~~\a}_{\n\r|}C_{\x}^{d\,\t}
  \nonumber \\
  &&+\frac{4}{9}\,C^{*e\a\x}{\widehat{T}}^{b\l}_{~~(\m|\s)}
  {\widehat{T}}^{c~~\s}_{\n\r|}C_{\x}^{d\,\t}\,
  \Big]\,.
\end{eqnarray}
It is impossible to get an expression with three ghosts, one $C^{*}$
and no field, by acting with $\d$ on $c_3\,$.  We 
can thus assume without loss of generality that $c_3$ vanishes,
which implies that  $(a_2,a_2)_{a.b.}$ should be $\g\,$-exact modulo total derivatives.

The use of the variable $\widetilde{T}_{\a\b}:=\ve^{\m\n}_{~~\a}\widehat{T}_{\m\n|\b}$
in place of $\widehat{T}_{\m\n|\r}(=-\frac{1}{2}\ve^{\a}_{~\m\n}\widetilde{T}_{\a\r})$ 
simplifies the calculations. We find, after expanding the products of $\ve$-densities,  
\begin{eqnarray}
        (a_2,a_2)_{a.b.}&=&\g\mu + d\n + f^a_{~bc}f_{ead}C^{*e\s\t}
  \Big[\,
    C^{b\m\a}{\widetilde{T}}^c_{\m\s}{\widetilde{T}}^d_{\a\t}    
  + C^{b\m}_{~~\s}{\widetilde{T}}^c_{\m\a}{\widetilde{T}}^{d\,\a}_{\;\t}
  \nonumber \\  
&&
 - \frac{2}{3}\,C^{b\m\a}{\widetilde{T}}^c_{\m\a}{\widetilde{T}}^d_{\s\t}   
  + C^{d\m}_{~~\s}{\widetilde{T}}^b_{\m\a}{\widetilde{T}}^{c\,\a}_{\;\t}  
  - \frac{1}{3}\,C^d_{\s\t}{\widetilde{T}}^{b\a\m}{\widetilde{T}}^c_{\a\m} 
  \Big]\,.
  \label{inter1}
\end{eqnarray}
We then use the only possible Schouten identity 
\begin{eqnarray}
        0 &\equiv& C^{*e\,\t}_{~[\s}C^{b\,\m}_{\;\a}{\widetilde{T}}^{c\,\s}_{\;\m}
        {\widetilde{T}}^{d\,\a}_{\;\t]}\nonumber \\
        &=& \frac{1}{24}\Big[
         - C^{*e\s\t}C^{b\m\a}{\widetilde{T}}^c_{\s\t}{\widetilde{T}}^d_{\m\a}
         + 2 C^{*e\s\t}C^{b\m\a}{\widetilde{T}}^c_{\s\m}{\widetilde{T}}^d_{\a\t}
   + 2 C^{*e\s\t}C^b_{\s\m}{\widetilde{T}}^c_{\t\a}{\widetilde{T}}^{d\,\a\m}
  \nonumber \\  
  && \qquad  - C^{*e\s\t}C^b_{\s\t}{\widetilde{T}}^c_{\m\n}{\widetilde{T}}^{d\,\m\n}
      - C^{*e\s\t}C^{b\m\n}{\widetilde{T}}^c_{\m\n}{\widetilde{T}}^d_{\s\t}
   + 2 C^{*e\s\t}C_{\s}^{b\,\m}{\widetilde{T}}^c_{\m\a}{\widetilde{T}}^{d\,\a}_{~\t}
        \Big] \label{Sca2a2}
\end{eqnarray}
in order to substitute in Eq.(\ref{inter1}) the expression of $C^{*e\s\t}C^{b\m\a}{\widetilde{T}}^c_{\m\s}{\widetilde{T}}^d_{\a\t}$ in terms of 
the other summands appearing in Eq.(\ref{Sca2a2}).  
Consequently, the following expression for $(a_2,a_2)_{a.b.}$ contains only linearly independent terms: 
\begin{eqnarray}
        (a_2,a_2)_{a.b.} &=&  \g\mu + d\n + C^{*e\s\t}\Big[
         {\textstyle\frac{1}{2}}f^a_{~bc}f_{dea}C^{b\m\a}{\widetilde{T}}^c_{\s\t}
   {\widetilde{T}}^d_{\m\a}+{\textstyle \frac{1}{6}}f^a_{~bc}f_{dea}C^{b\m\a}{\widetilde{T}}^d_{\s\t}
                {\widetilde{T}}^c_{\m\a}
         \nonumber \\
          &&+ {\textstyle{2}} f^a_{~c(b}f_{d)ea}C^{b\,\m}_{\s}{\widetilde{T}}^c_{\t\a}
        {\widetilde{T}}^{d\,\a}_{~\m}
        +{\textstyle \frac{1}{2}}f^a_{~b[c}f_{d]ea}C^{b}_{\s\t}{\widetilde{T}}^c_{\m\a}
        {\widetilde{T}}^{d\m\a}
        +{\textstyle \frac{1}{3}}f^a_{~bc}f_{dea}C^{d}_{\s\t}{\widetilde{T}}^c_{\m\a}
        {\widetilde{T}}^{b\m\a}
        \Big]\nonumber\,,
\end{eqnarray}
where we used that the structure constants of $\ca$ obey 
$f_{abc}\equiv \d_{ad}f^d_{~bc}=f_{[abc]}$. 

Therefore, the above expression is a $\g\,$-cobounday modulo $d$ if and only if 
 $f^a_{~bc}f_{dea}=0$, meaning that the internal algebra $\ca$ is nilpotent of order three. 
In turn, this implies \footnote{The internal metric $\d_{ab}$ being Euclidean, the 
condition $ f^a_{~bc}f_{aef}\equiv \d_{ad} f^a_{~bc}f^d_{~ef} = 0$ can be seen as 
expressing the vanishing of the norm of a vector in Euclidean space (fix $e=b$ and $f=c$), 
leading to $f^a_{~bc}=0$. } that $f^a_{~bc}=0$ and the deformation is trivial.


\subsection{Deformation in 5 dimensions}

Let us perform the same analysis for the candidate in five dimensions.

\subsubsection{First-order deformation}

First, $a_1$ must be computed from $a_2$ (given by (\ref{a2n5})), using the equation $\d a_2+\g a_1+d b_1=0\,$: 
\begin{eqnarray}
\d a_2&=&-3g^a_{~(bc)}\e^{\m\n\r\s\t}\pa_\l
h^{*\a\l}_{a~~\m}\pa_{[\n}C^b_{\r]\b}\pa_{\a[\s}C^{c\b}_{\t]}d^5
x\nonumber\\
&=&-d
b_1+3g^a_{~(bc)}\e^{\m\n\r\s\t}h^{*\a\l}_{a~~\m}[\pa_{\l[\n}C^b_{\r]\b}\pa_{\a[\s}C^{c\b}_{\t]}
+\pa_{[\n}C^b_{\r]\b}\pa_{\l\a[\s}C^{c\b}_{\t]}]d^5 x\,.
\nonumber \end{eqnarray}
We recall that it is a consequence of Theorem \ref{defa2} that $g^a_{~bc}$
is symmetric in its lower indices, thereby defining a commutative algebra. 
It is not an assumption, it comes from consistency.  
Therefore the first term between square bracket vanishes because of the symmetries of 
the structure constants $g^a_{~bc}$ of the internal commutative algebra $\cb\,$.  
We finally obtain, modulo trivial terms,
\begin{eqnarray}
a_1=\frac{3}{2}g^a_{~(bc)}\e^{\m\n\r\s\t}h^{*\a\l}_{a~~\,\m}\pa_{[\n}^{\ }C^{b\
\b}_{\r]}\left[\pa_{\b[\s}h^c_{\t]\l\a}-2\pa_{\l[\s}h^c_{\t]\a\b}\right]d^5
x\,.\nonumber \end{eqnarray}
The element $a_1$ gives the first-order deformation of the gauge transformations. 
By using the definition of the generalized de Wit--Freedman connections 
\cite{deWit:1979pe}, we get the following simple expression for $a_1$: 
\begin{eqnarray}
        a_1=g^a_{~(bc)}\e^{\m\n\r\s\t}h^{*\a\b}_{a~~\,\m}\pa_{[\n}^{\ }C^{b\ \l}_{\r]} 
        {\Gamma}^c_{\l[\s;\t]\a\b}  d^5x\,,
        \label{a1dWF}
\end{eqnarray}
where ${\Gamma}^c_{\l\s;\t\a\b}$ is the second spin-3 connection 
\begin{eqnarray}
        {\Gamma}^c_{\l\s;\t\a\b} =3\, \pa_{(\t}\pa_{\a}h^c_{\b)\l\s} 
        + \pa_{\l}\pa_{\s}h^c_{\t\a\b}   
- \frac{3}{2}\,\big(\pa_{\l}\pa_{(\t}h^c_{\a\b)\s}
        +\pa_{\s}\pa_{(\t}h^c_{\a\b)\l}\big)\nonumber
\end{eqnarray}
transforming under a gauge transformation $\d_{\l}h^a_{\m\n\r}=3\,\pa_{(\m}\l^a_{\n\r)}$ 
according to 
\begin{eqnarray}
        \d_{\l} {\Gamma}^c_{\r\s;\t\a\b} = 3\, \pa_{\t} \pa_{\a} \pa_{\b}\l^c_{\r\s}\,. \nonumber
\end{eqnarray}
The expression (\ref{a1dWF}) for $a_1$ implies that the deformed gauge transformations are 
\begin{eqnarray}
        \stackrel{(1)}{{\d_{\l}}} h^a_{\m\a\b} = 3\, \pa_{\m}\l^a_{\a\b} + g^a_{~(bc)} \,
        \e_{\m}^{~\,\n\r\s\t}\,{\Gamma}^b_{\g\n;\r\a\b}\,\pa_{\s}^{\ }\l^{c\ \g}_{\;\t}\,,
\label{defogt}
\end{eqnarray}
where the right-hand side must be completely symmetrized over the indices $(\m\a\b)\,$.       

The cubic deformation of the free Lagrangian, $a_0$, is obtained from $a_1$
by solving the top equation $\d a_1+\g a_0+d b_0=0$. 

Again, we consider the most general cubic
expression involving four derivatives and apply $\g$ to it, then we
compute $\d a_1$. We take the Euler-Lagrange derivative with respect to $C_{\a\b}$ of the sum of the two expressions, and multiply  by $C_{\a\b}$ to get  a sum of terms of the form $\ve C\pa^4 h \pa h$ or $\ve C\pa^3 h\pa^2 h$. These are not related by Schouten identities and are therefore independent; all coefficients of the obtained equation thus have  to vanish. When solving this system of equations,
we find that $g_{abc}\equiv \d_{ad}g^d_{~bc}$ must be completely symmetric. 
In other words, the corresponding internal commutative algebra $\cb$ 
possesses an invariant norm. As for the algebra $\ca$ of the $n=3$ case, 
the positivity of energy requirement
imposes a positive-definite internal metric with respect to which the norm is defined.  

Finally, we obtain the following solution for $a_0$:
\begin{eqnarray}\nonumber
a_0&=&\frac{3}{2}g_{(abc)}\e^{\m\n\r\s\t} \Big\{-\frac{1}{8}\pa_\m\Box
h^a_\n\pa_\r h^b_\s h^c_\t 
+\frac{1}{2}\pa^3_{\m\a\b} h^a_\n\pa_\r
h^{b\a\b}_\s h^c_\t
+\frac{1}{4}\pa_\m\Box h^{a\a\b}_\n\pa_\r
h^b_{\s\a\b}h^c_\t\nonumber \\
&&\qquad+\frac{3}{8}\pa_\m\Box h^a_\n\pa_\r h^{b\a\b}_\s h^c_{\t\a\b}
-\frac{1}{2}\pa_\m\Box h^{a\a\b}_\n\pa_\r h^b_{\s\a\g} h^{c\ \g}_{\t\b}
-\frac{1}{2}\pa^{3\a\b}_\m h^a_\n\pa_\r h^b_{\s\a\g} h^{c\ \g}_{\t\b}
\nonumber \\
&&\qquad -\frac{1}{2}\pa^{3\a\b}_\m h^a_{\n\a\g}\pa_\r h^b_\s h^{c\ \g}_{\t\b}
-\frac{1}{4}\pa^{3\a\b}_\m h^a_{\n\a\b}\pa_\r h^{b\g\d}_\s
h^c_{\t\g\d}
-\frac{1}{2}\pa^{3\a\b}_\m h^a_{\n\g\d}\pa_\r h^b_{\s\a\b}
h^{c\g\d}_\t
\nonumber \\
&&\qquad +\pa^{3\a\b}_\m h^a_{\n\b\g}\pa_\r h^{b\g\d}_\s h^c_{\t\a\d}
+\frac{1}{2}\pa^2_{\m\a}h_\n^{a\a\b}\pa_\r^{2\g}h^b_\s
h^c_{\t\b\g}
-\pa^2_{\m\a}h_{\n\b\g}^a\pa_{\r\d}^2h^{b\a\b}_\s
h^{c\g\d}_\t \Big\}d^5 x\,.\nonumber \end{eqnarray}

\subsubsection{Second-order deformation}

The next step is the equation at order 2 : $(W_1,W_1)_{a.b.}=-2sW_2$. In particular,  its antighost 2 component reads $(a_2,a_2)_{a.b.}=\g c_2+\pa_{\m}j^{\m}_2\label{a2a21}\,.$ The left-hand side is directly computed from Eq.(\ref{a2n5}) : 
\begin{eqnarray}  (a_2,a_2)_{a.b.} &=&-g^a_{bc}g_{dea}^{}\ve^{\bar{\m}\bar{\n}\bar{\r}\bar{\s}\bar{\t}}\ve_\m^{\ \,\n\r\s\t}\d^{(\m}_{\bar{\t}} \d^{\a)}_\d 
(4\pa_{\bar{\m}}C_{\bar{\n}}^{*d\g}\pa_{\g\bar{\r}}C_{\bar{\s}}^{e\d}+2\pa_{\g\bar{\m}}C_{\bar{\n}}^{*d\g}\pa_{\bar{\r}}C_{\bar{\s}}^{e\d} )
 \pa_\n^{\ }C_\r^{b\b}\pa_{\a\s}C_{\t\b}^{c}\nonumber \\
&=&  -12 g^a_{b[c}g_{d]ea}^{} C^{*b\a\b}\widehat{U}_{\ \a}^{c\ \g|\m\n}\widehat{U}_{\b\g}^{d\ \,|\r\s}\widehat{U}^e_{\m\n|\r\s}
+ \g c_2 + \pa_{\m}j^{\m}_2\,. \nonumber\end{eqnarray}

The first term appearing in the right-hand side of the
above equation is a nontrivial element of $H(\g\vert d)$ . 
Its vanishing implies that the structure constants $g_{(abc)}$ of the 
commutative invariant-normed algebra $\cb$ 
must obey the associativity relation $g_{~\;b[c}^a g_{d]ea}^{\ }=0$. As for the spin-2 deformation problem (see \cite{Boulanger:2000rq}, Sections 5.4 and 6), this means that, modulo 
redefinitions of the fields, there is no cross-interaction between different kinds of spin-3 gauge fields provided the internal metric in $\cb$ is positive-definite --- which is demanded 
by the positivity of energy. 
The cubic vertex $a_0$ can thus be written as a sum of independent self-interacting vertices, one for each field $h_{\m\n\r}^a\,$, $a=1,\ldots,N\,$. 
Without loss of generality, we may drop the internal index $a$ and consider only one \emph{single} self interacting spin-3 gauge field $h_{\m\n\r}\,$.  
 
\section{Conclusion}
\label{sec:concl}

By relaxing the parity requirement imposed in \cite{Bekaert:2005jf}, the present work 
completes the classification of the consistent non-Abelian perturbative 
deformations of Fronsdal's 
spin-3 gauge theory, under the assumptions of Poincar\'e invariance and locality. 

In \cite{Bekaert:2005jf} the first-order cubic vertex of Berends, Burgers, and van Dam
\cite{Berends:1984wp} was recovered. 
However, the latter vertex leads to inconsistencies when continued to second order in the 
coupling constant, as was shown in several places 
\cite{Bengtsson:1983bp,Berends:1984rq,Bengtsson:1986bz}. 
A new first-order non-Abelian deformation, leading to a cubic vertex, was also found in \cite{Bekaert:2005jf}.  
It is defined in space-time dimension $n\geqslant 5$ and passes the second-order 
test where the vertex of \cite{Berends:1984wp} shows an inconsistency.   

In the present paper, by explicitly breaking parity invariance, we obtained two more consistent non-Abelian first-order deformations, leading to a cubic vertex in the Lagrangian. 
The first one is defined in $n=3$ and involves a multiplet of gauge fields  
$h^a_{\m\n\r}$ taking values in an internal, anticommutative, invariant-normed 
algebra $\ca\,$. 
The second one lives in a space-time of dimension $n=5$, the fields taking value in an 
internal, commutative, invariant-normed algebra $\cb\,$. 
Taking the metrics which define the inner product in $\ca$ and $\cb$ positive-definite 
(which is required for the positivity of energy), the $n=3$ candidate gives rise to inconsistencies when continued to perturbation order two, whereas the $n=5$ one passes the 
 test and can be assumed to involve only \emph{one} kind of 
self-interacting spin-3 gauge field $h_{\m\n\r}$, bearing no internal ``color'' index.

Remarkably, the cubic vertex of the $n=5$ deformation is rather simple. 
Furthermore, the Abelian gauge transformations are deformed by the addition of a term involving the second de Wit--Freedman connection in a straightforward way, cf. 
Eq.(\ref{defogt}). 
The relevance of this second generalized Christoffel symbol in relation to a hypothetical spin-3 covariant derivative was already stressed in \cite{Bengtsson:1983bp}.     

It is interesting to compare the results of the present spin-3 analysis with those found 
in the spin-2 case first studied in \cite{Boulanger:2000ni}. 
There, two parity-breaking first-order consistent non-Abelian deformations of 
Fierz-Pauli theory were obtained, also living in dimensions $n=3$ and $n=5$.
The massless spin-2 fields in the first case bear a color index, the internal algebra 
$\widetilde{\ca}$ 
being commutative and further endowed with an invariant scalar product. 
In the second, $n=5$ case, the fields take value in an anticommutative, invariant-normed 
internal algebra $\widetilde{\cb}$. 
It was further shown in \cite{Boulanger:2000ni} that the $n=3$ first-order consistent 
deformation could be continued to \emph{all} orders in powers of the coupling constant, the 
resulting full interacting theory being explicitly written down \footnote{Since the 
deformation is consistent, starting from $n=3$ Fierz-Pauli, the complete $n=3$ 
interacting theory of \cite{Boulanger:2000ni} describes no propagating physical 
degree of freedom. 
On the contrary, the topologically massive theory in \cite{Deser:1981wh,Deser:1982vy} 
describes a massive graviton with \emph{one} propagating degree of freedom (and not 
\emph{two}, as 
was erroneously typed in \cite{Boulanger:2000ni}; N.B. wants to thank S. Deser for having  
pointed out this to him).}. 
However, it was not determined in \cite{Boulanger:2000ni} whether the $n=5$ candidate could 
be continued to all orders in the coupling constant. Very interestingly, 
this problem was later solved in \cite{Anco:2003pf}, where a consistency condition was 
obtained at second order in the deformation parameter, \emph{viz} the algebra 
$\widetilde{\cb}$ must be nilpotent of order three. 
Demanding positivity of energy and using  
the results of \cite{Boulanger:2000ni}, the latter nilpotency condition implies      
that there is actually no $n=5$ deformation at all: the structure constant of the 
internal algebra $\cb$ must vanish \cite{Anco:2003pf}. 
Stated differently, the $n=5$ first-order deformation 
candidate of \cite{Boulanger:2000ni} was shown to be inconsistent \cite{Anco:2003pf} when continued at second order in powers of the coupling constant, in analogy with the spin-3 
first-order deformation written in \cite{Berends:1984wp}.   

In the present spin-3 case, the situation is somehow the opposite.
Namely, it is the $n=3$ deformation which shows inconsistencies when going to second order, 
whereas the $n=5$ deformation passes the first test. Also, in the $n=3$ case the fields take 
values in an anticommutative, invariant-normed internal algebra $\ca$ whereas the fields 
in the $n=5$ case take value in a commutative, invariant-normed algebra $\cb\,$. 
However, the associativity condition deduced from a second-order consistency condition is 
obtained for the latter case, which implies that the algebra $\cb$ is a direct sum of 
one-dimensional ideals. 
We summarize the previous discussion in Table \ref{table1}. 
\begin{table}[!ht]
\centering
\begin{tabular}{|c||c|c|}
\hline 
 & $s=2$  & $s=3$ \\ \hline \hline
$n=3$& $\widetilde{\ca}$ commutative, invariant-normed 
&  $\ca$ anticommutative, invariant-normed and \\
 &  & nilpotent of order $3$ \\ \hline
$n=5$ & $\widetilde{\cb}$ anticommutative, invariant-normed and & $\cb$ commutative, invariant-normed and\\ 
&  nilpotent of order $3$ & associative \\
\hline
\end{tabular}
\caption{\it Internal algebras for the parity-breaking first-order 
deformations of spin-$2$ and spin-$3$ free gauge theories.\label{table1}}
\end{table}

It would be of course very interesting to investigate further the $n=5$ deformation 
exhibited here, since if the deformation can be consistently continued to all orders 
in powers of the coupling constant, this would give the first consistent interacting 
Lagrangian for a higher-spin gauge field.  
We hope to come back to this issue in the near future. 
 
\begin{acknowledgments}
We thank X. Bekaert and Ph. Spindel for comments and discussions. 
The work of S.C. is supported in part by the ``Interuniversity Attraction
Poles Programme -- Belgian Science Policy '', by IISN-Belgium
(convention 4.4505.86) and by the European Commission FP6
programme MRTN-CT-2004-005104, in which she is associated
with the V.U.Brussel (Belgium).
\end{acknowledgments}

\appendix*
\section{Schouten Identities}
\label{A}
%
The Schouten identities are identities due to the fact that in $n$ dimensions the antisymmetrization over any $n+1$ indices vanishes. These identities obviously depend on the dimension and relate functions of the fields.
As it is important to have a real basis when solving equations, this appendix is devoted to finding a basis for various sets of functions, depending on their structure and the number of dimensions. We consider Poincar\'e-invariant and parity-breaking functions, and we are interested mainly in the $\g$-nontrivial quantities. The fields may have an internal index but we only write it when necessary.
%
\subsection{Functions of the structure $\ve C^*\widehat{T}\widehat{T}$ in $n=4$}
\label{ectt}
%
In order to achieve the four-dimensional study of the algebra deformation in $D$-degree 2, a list of the Schouten identities is needed for the functions of the structure $\ve C^*\widehat{T}\widehat{T}$. 
The space of these functions is spanned by 
\begin{eqnarray}
T_1^{bc}=\ve^{\m\n\r\s}\
C^{*a\a}_{\m}\ \widehat{T}^{b~\ \b}_{\n\r\vert}\ \widehat{T}_{\s\a\vert \b}^c\;,\;
T_2^{bc}=\ve^{\m\n\r\s}\
C^{*a\a}_{\m}\ \widehat{T}^{b~\ \b}_{\n\r\vert}\ \widehat{T}_{\s\b\vert \a}^c\;,\;
T_3^{[bc]}=\ve^{\m\n\r\s}\ C^{*a\a\b}\
\widehat{T}^{b}_{\m\n\vert \a}\ \widehat{T}_{\r\s\vert \b}^c\,.
\nonumber \end{eqnarray}
There are two Schouten identities. Indeed, one should first notice that all Schouten identities are linear combinations of identities with the structure
$\d^{[\a\b\g\d\e]}_{[\m\n\r\s\t]}\ve^{\m\n\r\s} C^*\widehat{T}\widehat{T}=0\,,$ where the indices $\a\b\g\d\e\t$ are contracted with the indices of the ghosts and where $\d^{[\a\b\g\d\e]}_{[\m\n\r\s\t]}=\d^{[\a}_{[\m}\d^\b_\n\d^\g_\r\d^\d_\s\d^{\e]}_{\t]}\,.$ Furthermore, there are only two independent identities of this type:
\begin{eqnarray}
\d^{[\a\b\g\d\e]}_{[\m\n\r\s\t]}\ve^{\m\n\r\s} C^{*\t}_\a\widehat{T}^b_{\b\g\vert \l}\widehat{T}^{c~\l}_{\d\e\vert}=0\;,\;
\d^{[\a\b\g\d\e]}_{[\m\n\r\s\t]}\ve^{\m\n\r\s} C^{*\l}_\a\widehat{T}^{b~\,\t}_{\b\g\vert }\widehat{T}^{c}_{\d\e\vert\l}=0\,.
\nonumber \end{eqnarray}
Expanding the product of $\d$'s, one finds that the first identity implies that $T_1^{bc}$ is symmetric: $T_1^{bc}=T_1^{(bc)}\,,$
while the second one relates $T_2^{bc}$ and $T_3^{[bc]}\,$: $T_2^{bc}=T_3^{[bc]}\,.$

So, in four dimensions, a basis of the functions with the structure $\ve C^*\widehat{T}\widehat{T}$ is given by $T_1^{(bc)}$ and $T_3^{[bc]}\,.$

\subsection{Functions of the structure $\ve h^*\widehat{T}\widehat{U}$ in $n=4$}
\label{ehtu}
%
These functions appear in the study of the algebra deformation in $D$-degree 3, $n=4$ . 
They are completely generated by the following terms: 
\begin{eqnarray}
&T_1=\ve^{\m\n\r\s}h_{\ \m}^{*a\a\b}\widehat{T}^b_{\n\g\vert \b}\widehat{U}_{\r\s|\a}^{c\ \ \ \g} \;,\;
T_2=\ve^{\m\n\r\s}h_{\ \m}^{*a\a\b}\widehat{T}^b_{\n\b\vert \g}\widehat{U}_{\r\s|\a}^{c\ \ \ \g}\;,\;
T_3=\ve^{\m\n\r\s}h^{*a\a}\widehat{T}^{b~~\b}_{\m\n\vert }\widehat{U}_{\r\s|\a\b}^c\;,& \nonumber \\
&T_4=\ve^{\m\n\r\s}h^{*a}_{\m}\widehat{T}^{b\a\b\vert }_{\ ~~~ \n}\widehat{U}_{\r\s|\a\b}^c\;,\;
T_5=\ve^{\m\n\r\s}h_{\ \m}^{*a\a\b}\widehat{T}^b_{\n\r\vert \g}\widehat{U}_{\s\a|\b}^{c\ \ \ \g}\;.&\nonumber 
\end{eqnarray}

There are three Schouten identities:
\begin{eqnarray}
&\d^{[\a\b\g\d\e]}_{[\m\n\r\s\t]}\ve^{\m\n\r\s} h^{*~\,\t}_{\a\l}  \widehat{T}_{\b\g\vert \h}\widehat{U}_{\d\e\vert}^{~~\l\h}=0\;,\;
\d^{[\a\b\g\d\e]}_{[\m\n\r\s\t]}\ve^{\m\n\r\s} h^{*}_{\a}  \widehat{T}_{\b\g\vert \l}\widehat{U}_{\d\e\vert}^{~~\t\l}=0\;,&\nonumber \\
&\d^{[\a\b\g\d\e]}_{[\m\n\r\s\t]}\ve^{\m\n\r\s} h^{*~\,\l}_{\a\h}  \widehat{T}_{\b\g\vert \l}\widehat{U}_{\d\e\vert}^{~~\t\h}=0
\;.&\nonumber 
\end{eqnarray}
An explicit expansion of these identities yields the useful relations
$$T_3+2T_2+2T_5=0\;,\ T_3-T_4=0\;,\ T_1=0\;.$$
%
\subsection{Functions of the structure $\ve \, C^*\widehat{U}\widehat{U}$ in $n=4$}
\label{ecuu}
%
The Schouten identities for the functions of the structure $\ve \, C^*\widehat{U}\widehat{U}$ in $n=4$ are needed for the analysis of the algebra deformation in $D$-degree four.
The functions at hand are generated by $T_1^{[bc]}=\ve^{\m\n\r\s}\
C^{*a}_{\a\b}\ \widehat{U}^{b\ \ \a\g}_{\m\n|}\ \widehat{U}^{c\ \ \b}_{\r\s|\ \ \g}\,$ and $T_2^{bc}=\ve^{\m\n\r\s}\ C^{*a}_{\m\b}\ \widehat{U}^{b\ \ \a\g}_{\n\r|}\ \widehat{U}^{c\ \b}_{\ \s\ |\a\g}\;.$ However, these vanish because of the Schouten identities
\begin{eqnarray}
\d^{[\a\b\g\d\e]}_{[\m\n\r\s\t]}\ve^{\m\n\r\s}C^{*\l}_{\a}\widehat{U}^{b~~\t\h}_{\b\g\vert}\widehat{U}^{c}_{\g\d\vert\l\h}=0\;,\;
\d^{[\a\b\g\d\e]}_{[\m\n\r\s\t]}\ve^{\m\n\r\s}C^{*\t}_{\a}\widehat{U}^{b~~\l\h}_{\b\g\vert}\widehat{U}^{c}_{\g\d\vert\l\h}=0\;.
\nonumber \end{eqnarray}
Indeed, they imply that
$T_1^{[bc]}+T_2^{bc}=0$ and $T_2^{bc}=T_2^{(bc)}$, which can be satisfied only if $T_1^{[bc]}=T_2^{(bc)}=0\,.$

%
\subsection{Functions of the structure $\e C \pa^3 h h$ and $\e C \pa^2 h\pa  h$  in $n=3$}
\label{cpahh}
%
These functions appear when solving $\d a_1+\g a_0=db_0$ in Section \ref{dim3peu}. In generic dimension ($n>4$), there are respectively 45 and 130 independent functions in the sets $\e C \pa^3 h h$ and $\e C \pa^2 h\pa  h$ . In three dimensions, there are 108 Schouten identities relating them, which leave only 67 independent functions. We have computed all these identities and the relations between the 108 dependent functions and the 67 independent ones. However, given their numbers, they will not be reproduced here. 


\end{document}